\documentclass[]{sig-alternate}  

\usepackage{version}
\includeversion{techreport}
\excludeversion{ccs}

\makeatletter
\let\@copyrightspace\relax
\makeatother

\usepackage{endnotes}
\usepackage{url}
\usepackage{subfigure}
\usepackage{graphicx}
\usepackage{amssymb,amsmath}
\usepackage{algorithmic}
\usepackage{comment}
\usepackage{mdwlist}
\usepackage[usenames,dvipsnames]{color}
\usepackage{cite}
\usepackage{color,url,algorithm,algorithmic,comment,multirow,cite, mdwlist}   

\newcommand\mn[1]{} %

\newcommand{\matt}[1]{{\color{black}{#1}}} %
\newcommand{\nikita}[1]{{\color{black}{#1}}} 
\newcommand{\prateek}[1]{{\color{black}{#1}}} %

\newcommand{\paragraphb}[1]{\vspace{0.03in}\noindent{\bf #1} }

\begin{document}

\begin{ccs}
\conferenceinfo{CCS'11,} {October 17--21, 2011, Chicago, Illinois, USA.}
\CopyrightYear{2011}
\crdata{978-1-4503-0948-6/11/10}
\clubpenalty=10000
\widowpenalty = 10000
\end{ccs}

\title{Stealthy Traffic Analysis of Low-Latency Anonymous Communication Using Throughput Fingerprinting}

\numberofauthors{1} 
\author{\alignauthor Prateek Mittal, Ahmed Khurshid, Joshua Juen, Matthew Caesar, Nikita Borisov \\
\affaddr{University of Illinois at Urbana-Champaign} \\
\email{\{mittal2, khurshi1, juen1, caesar, nikita\}@illinois.edu} \\
} 

\maketitle

\begin{abstract}

Anonymity systems such as Tor aim to enable users to communicate in a manner
that is untraceable by adversaries that control a small number of machines.
\mn{What about relays that are throttled in tor config? Our attack should still work, no?} To provide efficient service to users, these anonymity systems make full
use of forwarding capacity when sending traffic between intermediate relays.
In this paper, we show that  doing this leaks information about the set of 
Tor relays in a circuit (path).
We present attacks that, with high confidence and based solely on throughput information, can 
(a) \prateek{reduce the attacker's uncertainty about} the bottleneck relay of any Tor circuit whose throughput can be observed, 
(b) exactly identify the guard relay\nikita{(s)} of a Tor user when circuit throughput can be observed over multiple connections, and %
(c) identify \nikita{whether two concurrent TCP connections belong to the same Tor user, breaking unlinkability}. 
Our attacks are stealthy, and cannot be \nikita{readily} detected by a user or by Tor relays. 
We validate our attacks using experiments over the live Tor network. We find that 
the attacker can substantially reduce the \prateek{entropy of a bottleneck relay distribution} of a Tor 
circuit whose throughput can be observed---the entropy gets reduced by a factor of 
$2$ in the median case. \mn{Factor of 2 is weird when talking about entropy} Such information leaks from a single Tor circuit can be 
combined over multiple connections %
to exactly identify a user's guard relay\nikita{(s)}.  
Finally, we are also able to link two connections from the same initiator with a crossover 
error rate of less than $1.5\%$ in under $5$ minutes.  Our attacks are \nikita{also} more accurate and 
require fewer resources than previous attacks on Tor.

\end{abstract}

\begin{ccs}
\end{ccs}

\section{Introduction}
\label{sec:intro}

With the advent of sophisticated monitoring                                     	
technologies~\cite{netflow,dlp-symantec}
 coupled with coordination within
governments and across businesses to exchange and mine information,
communication on the Internet is increasingly becoming less private.
At the same time, advancing use and pervasiveness of networks have increased
the danger of misuse of that information, leading to potential
invasions of user privacy, and fear of retribution against whistleblowers,
unofficial leaks, and political activists.
To counter this threat, {\em anonymous communication systems} such as
Babel~\cite{Babel}, Mixmaster~\cite{mixmaster}, and Mixminion~\cite{danezis:oakland03} improve anonymity by
routing communication through an overlay network that masks identities of the
endpoints. However, these systems work by introducing large and variable
latencies between the endpoints, making them unsuitable for interactive applications.
To fill this need, low-latency anonymous communication
services like AN.ON~\cite{an.on}, Onion Routing~\cite{onion-routing-service}, Freedom~\cite{freedom}, I2P~\cite{i2p:03} and 
Tor~\cite{dingledine:sec04} %
were developed, which forward packets directly over
low-delay circuits.  Tor has achieved great success, servicing hundreds of
thousands of users and carrying terabytes of traffic daily~\cite{metrics-portal}, and is currently
being used as a crucial service by end users, government activists,
journalists, as well as protecting business and military communications~\cite{who-uses-tor}.  %

Low-latency anonymous communication networks are more vulnerable to a global 
passive adversary than other high-latency communication 
designs~\cite{chaum:cacm81,mixmaster,danezis:sec03} because timing attacks can be 
used to link relayed copies of a communication stream~\cite{syverson:pet00}.
Most users, however, are concerned about less powerful, \emph{partial}
adversaries who can observe (or even compromise) only a fraction of the
network.  A typical goal for low-latency systems is to ensure that an adversary
who observes or controls a fraction $f$ of the network can compromise the
anonymity of $f^2$ of the flows. The hope is that making $f$ large would
require a high resource expenditure, especially as networks grow large (e.g.,
the Tor network had more than $2\,000$ active relays with an aggregate bandwidth of
4\,Gbps as of May 2011\footnote{\newcounter{torstatusfn}\setcounter{torstatusfn}{\value{footnote}}\url{http://torstatus.blutmagie.de/}}).

Further research has demonstrated that it is not, in fact, necessary to directly observe the 
network to learn where a flow is being forwarded. Instead, probes can be used to test a flow's 
presence at a particular location in the network by creating an observable interference 
pattern~\cite{murdoch:oakland05,evans:usenix09,chakravarty:esorics10}, 
 or by otherwise correlating characteristics seen in the anonymous flow with a particular location~\cite{hopper+:tissec10}.  
In order to fully understand the risks of using a given anonymity network, it is necessary to 
understand what types of attacks of this class are possible and what resources are required for a 
certain probability of success.  Such understanding will let users make informed security decisions; 
for example, a ``low-resource'' congestion attack on the Tor network originally proposed by Murdoch 
and Danezis in 2005~\cite{murdoch:oakland05} is no longer considered practical for the current size 
of the network without new techniques to improve its scalability~\cite{evans:usenix09}.
Users may also find comfort in the fact that the state-of-the-art attacks could 
be detected by Tor clients and relays, as they require the 
insertion of interference patterns or malicious content in flows.

In contrast, we examine a new collection of \emph{stealthy} attacks on the Tor network that are 
based on detailed observations of the \emph{throughput} of an anonymous flow.  
\mn{N: I rewrote this sentence} We called our attacks stealthy because, although they sometimes perform active \emph{measurements}, they
do so by acting as completely ordinary Tor clients, and therefore Tor users and relays cannot
reliably detect that an attack is in progress.
We observe that the 
throughput of a Tor flow can be used as a fingerprint of the bottleneck relay \nikita{(i.e., the relay with
minimal forwarding capacity in the flow path)} used to forward it.  
By observing the dynamics of the throughput, it is also possible to identify when two flows share the same set 
(or subset) of relays. We note that in contrast to prior work, our attacks do not require the adversary to insert interference patterns 
or other malicious content in flows, 
and are thus not observable by a user/relay. We also perform a detailed study of how the throughput of the anonymous flow 
is influenced by scheduling, flow-control, and congestion-control algorithms at both the transport 
(i.e., TCP) and application (i.e., Tor) layer.  We use the effects of these algorithms to design a 
passive attack that can show whether two flows share the same circuit and thus originate from the same user. 
To the best of our knowledge, this is the first study of flow throughput based attacks on Tor. 
Overall, our study demonstrates attacks that use threat models and resource requirements that 
differ significantly from attacks using other flow features.  It also highlights the complexity of 
designing anonymous communication systems by demonstrating the impact that architectural and 
implementation decisions can have on the security of the system. 

\subsection{Overview of Attacks}

\paragraphb{Information leakage via circuit throughput:} 
We show that it is possible to exploit heterogeneity of throughput of Tor relays to
learn information about which relays used to forward a given stream.  In particular,
two circuits that share the same bottleneck relay will have highly correlated throughput.  Thus,
by forming one-hop probe circuits through all of the Tor relays, it is possible to gain some (probabilistic) information
about the bottleneck relay of any Tor stream whose throughput can be observed.  This observation also 
reveals some information about other relays forwarding the stream, since their observed one-hop throughput
must be higher than the bottleneck.  We performed our attacks on a subset of the live Tor network, and found that we were 
able to effectively reduce the \prateek{entropy of the bottleneck relay distribution}. 
\mn{What does ``effectively'' mean here?}
Using this information, we were able to uncover the identities of the guard relay (the first relay in the path) over 
multiple connection attempts. \prateek{We note that while  learning a user's relays is not sufficient by 
itself to compromise the user identity in Tor, our attacks serve as a stepping stone for completely de-anonymizing 
the user; for example, several attacks in the literature rely on a user's guard relays being known to the 
adversary~\cite{hopper:ccs07,hopper+:tissec10,chakravarty:esorics10}.}
We \matt{also} show that our attacks can also be applied to de-anonymize location 
hidden services that act as Tor relays.

\paragraphb{Stream linkability attack:} 
We show that throughput characteristics of two streams that are 
multiplexed over the same circuit have a unique characteristic: throughput of each stream
repeatedly drops to zero during mutually exclusive periods of time, leading to a strong negative correlation.
Thus if a client is communicating over Tor with two servers simultaneously, and if the servers collude, 
then they can reliably infer that they are communicating with the same client. Our experimental
results indicate that after roughly 5 minutes of observations the attacker could reliably 
(with a crossover error rate of less than 1.5\%) identify if the streams have been multiplexed over a common 
circuit. This is a significant improvement of the latency-based stream linking attack by Hopper et al.~\cite{hopper:ccs07},
which had a crossover error rate of 17\% and required malicious content to be inserted into web pages. 

We have validated all of our attacks using experiments over the live Tor network.  \nikita{Our source 
code and analysis scripts are available at} our project webpage~\cite{project-webpage}. 

\textbf{Roadmap.} The rest of the paper is organized as follows. We present an overview of low-latency anonymity systems
and discuss related attacks in Section~\ref{sec:related}. In Section~\ref{sec:circuit-throughput} we show that there is a high degree of %
heterogeneity in observed circuit throughput in the Tor network, 
and that circuit throughput can be used to reduce the entropy of the corresponding Tor relays.
We also present several
\emph{statistical disclosure attacks} 
based on 
information leakage from circuit throughput to uncover the identities of guard relays and relays that offer location-hidden services. 
We present our passive stream linkability 
attack in Section~\ref{sec:stream-throughput}. %
Finally, we discuss the implications of our attacks in 
Section~\ref{sec:discussion} %
and conclude in Section~\ref{sec:conclusion}.

\section{Background and Related Work}
\label{sec:related}

We start by discussing work on designing low-latency anonymous
communication systems.  We then describe three key attacks on these systems:
side channel attacks, latency attacks, and long-term attacks. %

\subsection{Tor Background}

Tor~\cite{dingledine:sec04} is a popular low-latency anonymous communication system. 
Deployed in 2003, Tor now serves hundreds of thousands of users and carries 
terabytes of traffic daily.
As of May 2011, the network comprises more than $2\,000$ relays.\footnotemark[\value{torstatusfn}]
Users (clients) download a list of relays from central directory authorities and build anonymous 
paths (called \emph{circuits}) using onion routing~\cite{syverson:pet00}. Tor clients build 
three-hop circuits for anonymous communication, where the three relays are respectively known 
as \emph{guard}, \nikita{\emph{middle}}, and \emph{exit} relays. The guard relay is always chosen 
from a fixed set of three relays that is unique to each client, to prevent certain 
long term attacks on anonymous communication~\cite{wright:ndss02,wright:oakland03}.%
In order to balance the load on the network, clients 
select Tor relays in proportion to their bandwidths (subject to certain constraints). %
A Tor client can multiplex individual TCP connections (called \emph{streams}) over a single 
Tor circuit. The lifetime of a Tor circuit is generally set to $10$ minutes. Finally, in 
addition to anonymous communication, Tor also provides support for \emph{location-hidden services}; 
clients can connect to these services without knowing their network identity. We refer the 
interested reader to \cite{dingledine:sec04} for a more detailed description of Tor.

\subsection{Tor Security}

Traditional security analyses of Tor~\cite{dingledine:sec04} assume that a user who controls (or observes) a
fraction $f$ of the network can compromise the anonymity of $f^2$ of all tunnels by end-to-end timing analysis 
(by observing the entry and exit point of a stream). 
Note that due to bandwidth-weighted relay selection, $f$ is best thought of as the fraction 
of Tor bandwidth controlled or observed by an adversary.  
This simple model, however, abstracts away many important
properties of the system that affect anonymity; recent research has shown that, when these properties are
properly considered, the potential for anonymity compromise is significantly greater than predicted by the model.
Some examples of such properties include the reliability of relays~\cite{borisov:ccs07}, the skew of the 
internal system clocks~\cite{murdoch:ccs06} and the topology of the underlying Internet paths used to forward
traffic between relays~\cite{feamster:wpes04,murdoch:pet07,edman-syverson:ccs09}.  A
particular class of attacks uses side-channel information to determine whether an anonymous flow is forwarded by a particular
Tor relay~\cite{murdoch:oakland05,back:ih01,evans:usenix09,hopper:ccs07,chakravarty:esorics10}; this effectively
increases the fraction of the network that is observed ($f$) with a moderate resource expenditure.  As our 
attacks fall into the same class, we next discuss these attacks in detail.

\subsection{Circuit Clogging Attacks}

Murdoch and Danezis~\cite{murdoch:oakland05} proposed an attack on the Tor network 
where an adversary aims to identify the Tor relays of a circuit by \emph{modulating} the 
sending rate of traffic through a circuit, and studying the effect of the modulations 
on the \emph{latency} characteristics (queueing delay) of individual Tor relays.  
A high correlation between the two indicates that the Tor relay is likely part of 
the original circuit. The attacks were tested in 2005, when the Tor network
comprised only a handful of relays. Since 2005, the Tor network has grown in size by 
two orders of magnitude and Evans et al.~\cite{evans:usenix09} recently showed that 
the Murdoch--Danezis attack no longer works on the Tor network.

Evans et al.~\cite{evans:usenix09} proposed a variant of the Murdoch--Danezis attack where 
an adversary sets up a very long path through the Tor network, consisting of loops, such that 
the additional traffic introduced by an adversary gets amplified (in order to effectively 
congest a relay). There exist simple countermeasures to such an amplification attack, like 
imposing a bound on the maximum circuit length in Tor. Houmansadr and Borisov~\cite{houmansadr:ndss11} 
also show that attempts to circumvent the bound by having the flow exit the Tor network and 
then re-enter it (by the means of forwarding proxies) can be mitigated using watermarking 
schemes. Note that without the ability to amplify traffic, significant resources are required 
in order to congest a relay.  

Chakravarty et al.~\cite{chakravarty:esorics10} proposed another variant of the Murdoch--Danezis attack 
in which an attacker modulates traffic rates of a circuit, and uses estimation tools to observe changes 
in forwarding capacity of Tor relays. 
In theory, their approach also allows the identification of Internet links used to carry an anonymous flow, with the 
potential to identify the identity of the Tor client. The attack, however, requires an extensive
infrastructure of vantage points and maps of Internet %
topologies to be fully effective. Furthermore, the 
experimental evaluation of the attack showed only moderate success rates even under controlled settings; \nikita{we expect} this 
is because modulation of circuit rates does not have any impact on the forwarding capacity of nodes/links 
that are already operating at full capacity (such as some Tor relays).  

Note that all of these attacks rely on coarse-grained effects of active modulation of circuit traffic. 
One mitigation for this attack approach is to reduce or eliminate the ability of two circuits to interfere with each 
other's performance by altering the packet scheduling algorithms used by Tor~\cite{mclachlan:fc08}.

\subsection{Latency Observations}

Hopper et al.~\cite{hopper:ccs07,hopper+:tissec10} studied information leaks in Tor that 
arise due to heterogeneous network latency. By measuring the round-trip times through a 
Tor circuit (using malicious JavaScript content inserted into web pages), it is possible to compute the likelihood of it following a particular path 
through the Tor network and eliminate some paths from consideration entirely. Additionally, it may be
possible to learn information about the approximate geographic location of the client. Some of our attack scenarios
are similar in spirit, but our attacks focus on throughput, rather than latency; in the stream linkability attack, we find
that this results in significantly better accuracy.%

A major limitation of circuit-clogging and latency attacks is %
that they are not stealthy; the traffic modulations and malicious content insertions
can easily be detected by clients/Tor relays. In contrast, our attacks use finer-grained information about 
\prateek{circuit throughput} that can be obtained by a simple download through Tor, without actively modulating circuit traffic. Thus, 
our attacks are \emph{stealthy} and cannot be detected by Tor relays and clients. 
Our techniques also enlarge the scope of possible threat models under which an attacker can perform traffic 
analysis. For example, prior attacks are only applicable to the scenarios where either the destination 
web server or the exit relay is compromised. In contrast, our 
attacks can be carried out by ISPs or other local traffic observers. %
Finally, the finer-grained information also allows us to achieve better accuracy while using fewer resources.

\subsection{Long-Term Anonymity}

If multiple communication rounds can be linked (e.g., if a user logs in frequently with a given pseudonym), the anonymity guarantees
provided will degrade over time.  For example, using the $f^2$ probability of circuit compromise, over a number of path reformulations,
the chance that at least one tunnel will be observed at both the entry and exit approaches one~\cite{wright:ndss02,wright:oakland03}.%
This observation, along with attacks that can force reformulations to occur~\nikita{\cite{overlier-syverson:oakland06,abbott+:pets07,borisov:ccs07}}, motivates the use of guard
relays in Tor, where each client uses a small fixed set of entry points into the network.  

A similar degradation can be observed in mix-based systems, where observations of the set of 
active users~\cite{berthold:cfp00,raymond:pet00} or senders and recipients~\cite{kesdogan:ih02} 
at each communication round allows the use of \emph{intersection attacks} to narrow down the set of 
potentially communicating parties.  A statistical version of this attack (\emph{statistical 
disclosure attack}) was proposed by 
Danezis~\cite{danezis:sec03}. %
Similarly, information learned by observing the throughput of circuits
can be used for a statistical disclosure attack, as we discuss in Section~\ref{sec:attacks}.

\section{Information Leakage via \\ Circuit Throughput}
\label{sec:attacks}
\label{sec:circuit-throughput}
\label{sec:circuit-attacks}

We next describe attacks on Tor that determine path information about a pair
of circuits
by correlating their observed throughput.
We first motivate our approach by noting that the capacities of Tor relays are quite
heterogeneous (Section~\ref{sec:obsvhet}).
We then present an attack to determine if two circuits share a common sub-path (Section~\ref{sec:subpath}),
and an attack to narrow in on which relays a given circuit traverses
(Section~\ref{sec:spechop}). We combine these probabilistic observations over multiple 
circuit reformulations to de-anonymize guard relays %
(Section~\ref{sec:guards}). %
Finally, we show the applicability of our attack on interactive traffic (Section~\ref{sec:interactive}).

\subsection{\matt{Heterogeneity in the Tor Network}}  %
\label{sec:obsvhet}

\begin{figure}[t]
\begin{center}
\includegraphics[width=1.7in,angle=270]{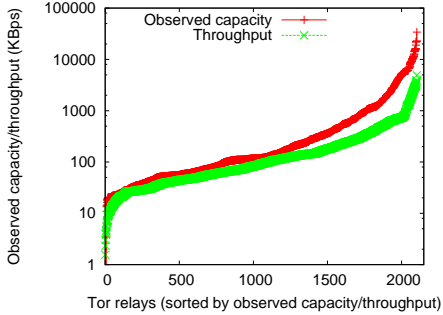} 
\end{center}
\vspace{-0.2in}
\caption{Observed capacity/throughput heterogeneity in the Tor network.}
\label{fig:capacity-throughput-relays}
\end{figure}

One fundamental observation used in our attacks is that the Tor network is 
composed of relays that have heterogeneous capacities. Figure~\ref{fig:capacity-throughput-relays} 
plots the advertised capacities of 2\,104 Tor relays, collected from the Tor directory service.  \mn{Aren't these TorFlow capacities?}
These capacity values are estimates based on the minimum of the maximum input or output bandwidth sustained 
over any ten second period at a particular Tor relay. This value is re-evaluated each day 
by every Tor relay and reported to the directory service. This information can be fetched using the 
\emph{getinfo desc/all-recent} command. By analyzing these capacity values 
we can see that there is a significant degree of heterogeneity in the capacities of the Tor relays.

From Figure~\ref{fig:capacity-throughput-relays}, we can see that around $38.6\%$ of these Tor relays 
have capacity (plotted on a log scale) below $100$\,KBps. This increases the probability of the Tor network 
being the bottleneck for most of the flows passing through Tor and makes it vulnerable to our attacks.      

In order to balance the load on the network,  Tor clients select relays in proportion to their capacities. 
Let $C_i$ denote the capacity of relay $i$. If there are $X$ circuits being constructed in the 
network, then the expected number of circuits that choose any particular relay $i$ is given by 
$X_i = 3 \cdot X \frac{C_i}{\Sigma_{i} C_i}$ (there is a factor of 3 because the Tor path consists of 3 relays). 
The capacity of relay $i$ should be distributed evenly 
amongst circuits that traverse it, given by $\frac{C_i}{X_i} = \frac{\Sigma_{i} C_i}{3 \cdot X}$. Thus we 
can see that in an ideal scenario, the throughput received by all circuits should be homogeneous.\footnote{This is a slight simplification; some relays cannot be in the first or third position, and those that can are underweighted when selecting the middle relay.}

Unfortunately, this is not the situation in the current Tor network. In spite of
load balancing, there exists a large range of 
throughput that individual relays can sustain, as shown in Figure~\ref{fig:capacity-throughput-relays}. 
To measure this, we individually probed the throughput of each available
Tor relay using a 3-hop circuit with our own guard and exit relay.
We used the stable version of Tor (Version 0.2.1.26, as of November 2010) to do the probing. 
We used two custom-built TCP client and server programs (details of these programs are presented in Section~\ref{sec:subpath}) 
to set up a flow through the relay being probed (using it as the middle relay). We ran both the client 
and the server programs at the same machine where the server sent data to the client as fast as possible (only limited by TCP's 
congestion control mechanism). We measured the throughput of the flow as observed by the client. %
Each relay was probed for $15$ seconds, and we computed the average observed throughput over the last $10$ seconds to avoid TCP's slow-start effect. 
As we probed each relay one by one, the entire experiment took about 
$18$ hours to complete. Note that we chose a high capacity vantage point to perform this experiment so that it did not 
become the bottleneck. We were able to probe 2\,104 out of 2\,429 relays listed in the Tor network consensus on July 24, 2011. The remaining 
325 relays did not respond to our circuit setup requests as they may have been offline when they were probed during our experiment.

There are three key factors that determine the throughput of a circuit: a) the capacity of the bottleneck relay, 
b) the number of active TCP flows between the bottleneck relay and the next hops and c) the number of other active circuits 
multiplexed over the TCP connection that carries the circuit in question\footnote{This formula makes the simplifying assumption that TCP results in fair sharing of bandwidth
between flows.  In reality, TCP is only RTT-fair~\cite{padhye+:ccr98}; this difference, however, has minimal impact on our analysis.}:\mn{How true is this for 
relays that cap their own bandwidth in tor-config?}

\begin{equation*}                    
  \parbox{0.8in}{\raggedright throughput of \\circuit $i$} = \frac{\text{bottleneck relay bandwidth}}{\text{\# of TCP flows} \times \parbox{0.9in}{\raggedleft \# of circuits in \\$i$'s TCP flow}}
\end{equation*}

There are several reasons for the lack of perfect load balancing in the Tor network: a) the first hop 
relay is chosen from a fixed set of three guard relays, b) the last hop is chosen 
from the set of relays whose exit policy meets the client's requirements, c) the peak capacity 
advertised by a relay is capped to an arbitrary threshold, d) the clients are not privy to current 
levels of traffic in the network, and hence select relays in a fully decentralized fashion, and e) 
the clients cannot predict stream characteristics in advance when selecting circuits for anonymous communication. 
We note that most of these reasons are fundamental limitations for any secure \matt{low-latency} anonymity system. 

\subsection{Circuit-Based Bandwidth Fingerprinting}
\label{sec:subpath}

In this section, we describe an attack that determines whether two circuits share a common sub-path by ``fingerprinting'' their communication. We do this by monitoring their throughput %
and using a simple statistical test to determine if their throughput is correlated.
\matt{We consider three representative experimental scenarios:}

\begin{figure*}[t]
\centering
\subfigure[]{{\fbox{\includegraphics[width=2.0in]{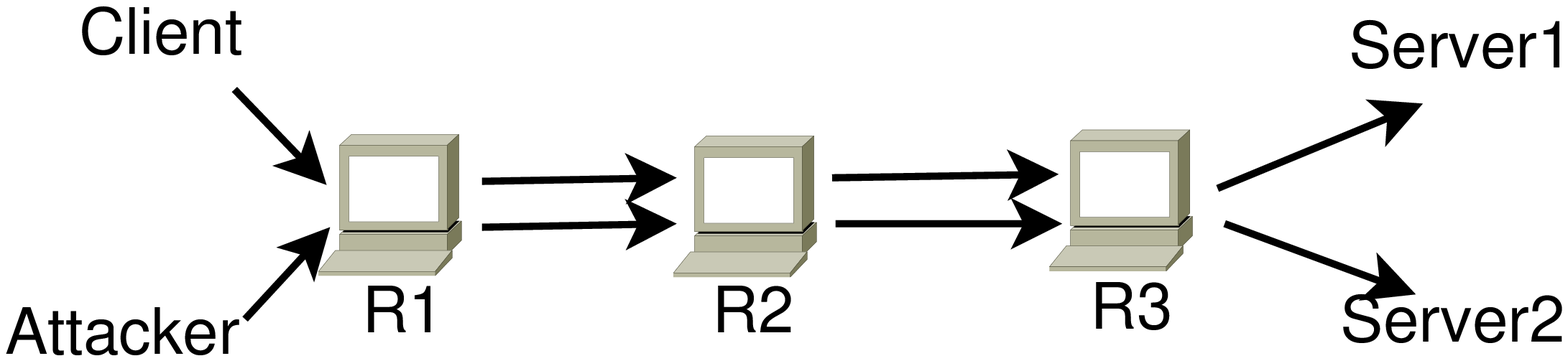}}}}
\subfigure[]{{\fbox{\includegraphics[width=2.0in]{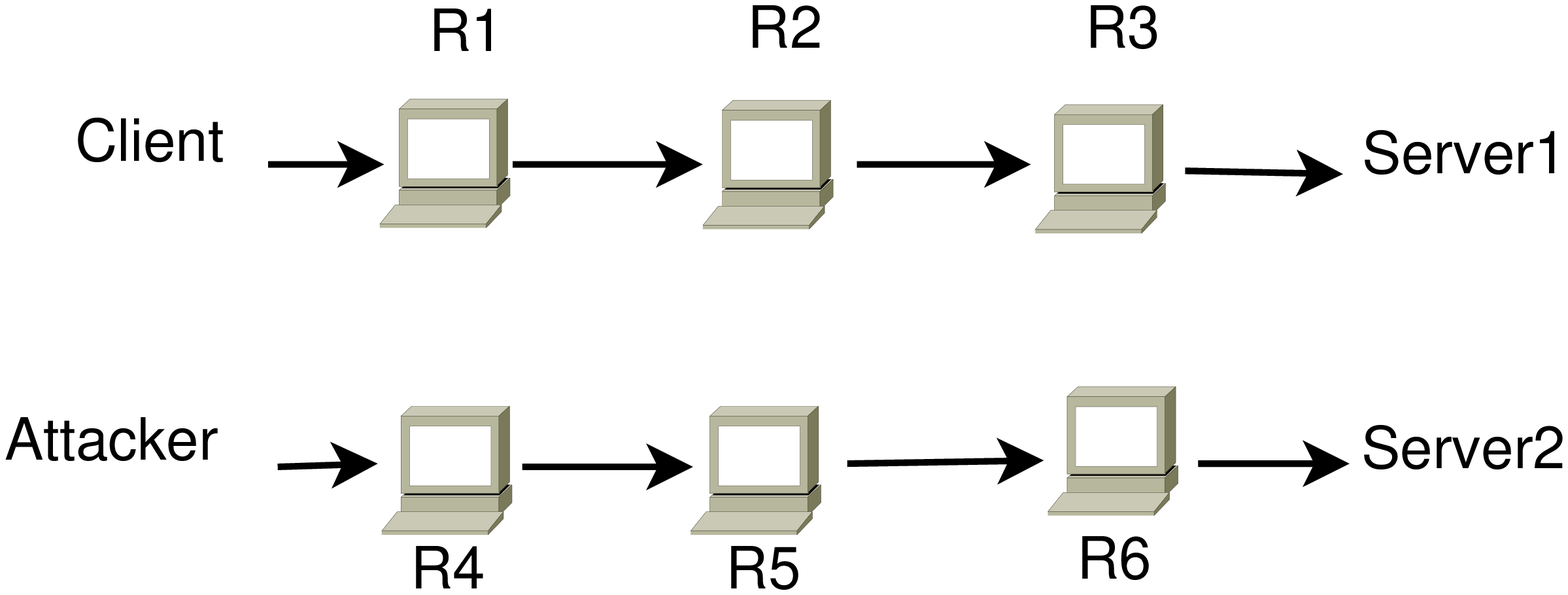}}}}
\subfigure[]{{\fbox{\includegraphics[width=2.0in]{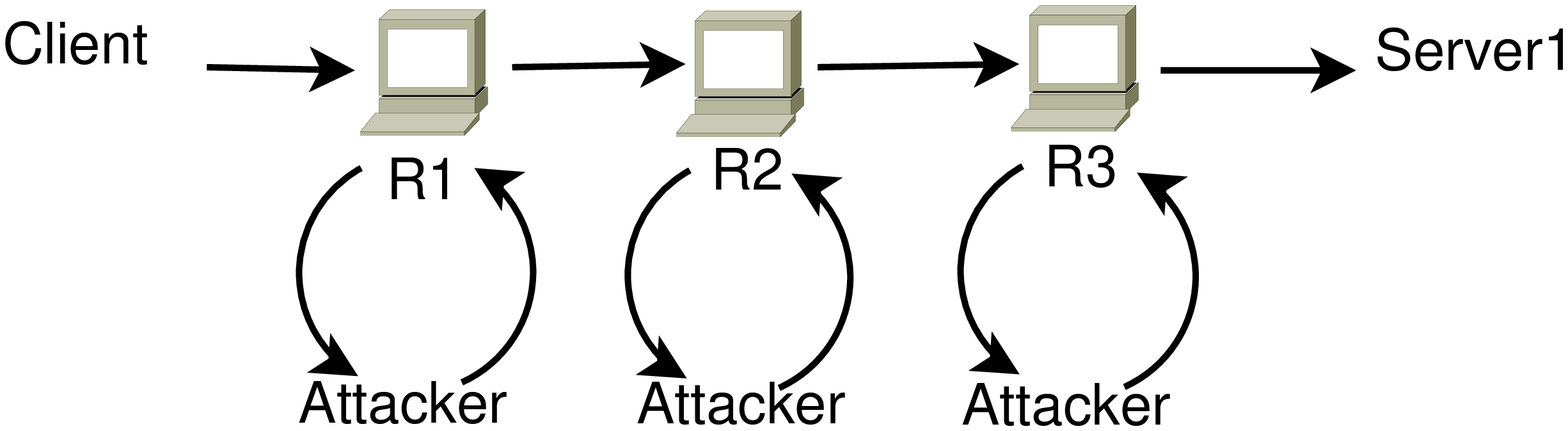}}}}
\caption{
Experimental scenarios for circuit-based throughput fingerprinting:
(a) {\em  All-Common:} Two circuits sharing all three Tor relays have similar throughput characteristics.
(b) {\em  None-Common:} Two circuits without any common Tor relays do not have  any similarity in their throughput characteristics.
(c) {\em  One-Common:} Two circuits with one or two common Tor relays will have similar throughput characteristics but only for the duration of time in which the  common Tor relays are the bottleneck relays in the circuit. 
}
\label{fig:scenarios}
\end{figure*}

\matt{{\em All-Common} (Figure~\ref{fig:scenarios}(a)):} If two Tor circuits have all relays in common, then the circuit throughput will be highly correlated. Since all 
three relays are common in both the circuits, any variation in the number of TCP flows or forwarding capacity at a relay will affect the throughput of both circuits 
in a similar fashion. %

\matt{{\em None-Common} (Figure~\ref{fig:scenarios}(b)):}
If two Tor circuits do not have any common relays, then the circuit throughput will not be correlated. 

\matt{{\em One-Common} (Figure~\ref{fig:scenarios}(c)):}
\matt{If two Tor circuits have at least one relay in common, but do not have all
relays in common, then there are two cases. 
If the shared relay is the bottleneck in both the circuits (\emph{One-Common-a}), then
the circuit throughput will be highly correlated.
If the shared relay is not the bottleneck in both the circuits (\emph{One-Common-b}),
then changes}
in its number of TCP flows or forwarding capacity will not affect the throughput of the two circuits, and thus their throughput will depend 
on their respective bottleneck relays. 

We performed experiments on the live Tor network to verify these observations. 
We used two client nodes and two server nodes, located in four separate geographical regions. 
On the client machines, we ran the stable version of Tor as of November 2010 (Version 0.2.1.26). We 
used the default path selection in the Tor clients, but turned off the use of entry guards (by setting \texttt{UseEntryGuards=0}), 
so that our clients sample circuits from the space of all possible circuits. For our 
experiments with all three common relays, we disabled preemptive 
circuit setup on the Tor client at the attacker's machine (using \texttt{\_\_DisablePredictedCircuits=1}). 
We used two custom-built TCP programs to work as the client 
and server in all of our experiments. The TCP client is a C program that connects with the TCP server, reads data sent by the server and 
computes the throughput of the flow periodically after a specific measurement interval. It uses the Tor SOCKS~\cite{SOCKS} interface
to perform download via Tor. The TCP server is a multithreaded server written in C that 
waits for connection requests from clients and once connected sends random bytes to the client as fast as possible. The sending rate of 
the server is only limited by the TCP congestion control algorithm. For simplicity of presentation, our experiments in this 
section assume that a user is interested in bulk data transfer. Our attacks are also applicable to interactive traffic; we demonstrate 
this in Section~\ref{sec:interactive}.                                                             

For each experiment, we set up a stream at the honest client, recorded the circuit 
that the Tor client assigned to that stream, and built the same circuit at the attacker's client. 
We used an upper bound of 30 seconds as the synchronization delay, and after this delay, 
we set up a new stream at the attacker's client.

\begin{figure*}[t]
\centering
\subfigure[]{{{\includegraphics[width=2.3in]{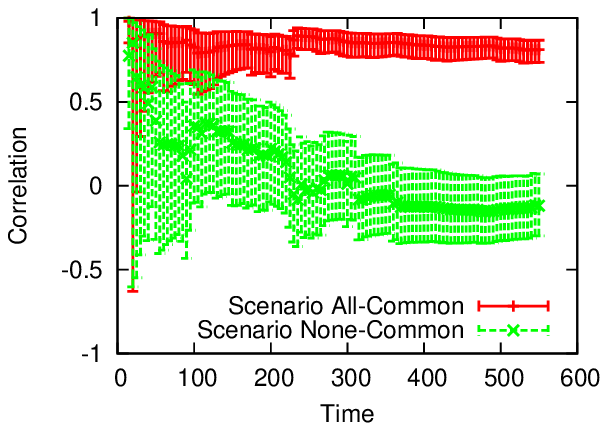}}}}
\hspace{0.7in}
\subfigure[]{{{\includegraphics[width=2.3in]{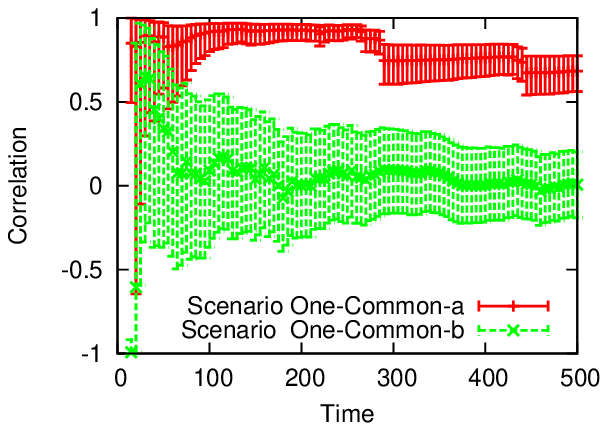}}}}
\caption{
(a) When the two circuits share all three relays (\emph{All-Common}), the correlation  is
very high. When the circuits share no common relays (\emph{None-Common}), the      correlation
is very low.
(b) When the two circuits share a single relay, and that relay is the bottleneck on both paths
(\emph{One-Common-a}), then the correlation is high.
If that relay is not the bottleneck on both paths (\emph{One-Common-b}), then the  correlation is low.
}
\label{fig:correlation-time}
\end{figure*}

\begin{figure*}[t]
\centering
\subfigure[c]{{{\includegraphics[width=1.5in,angle=270]{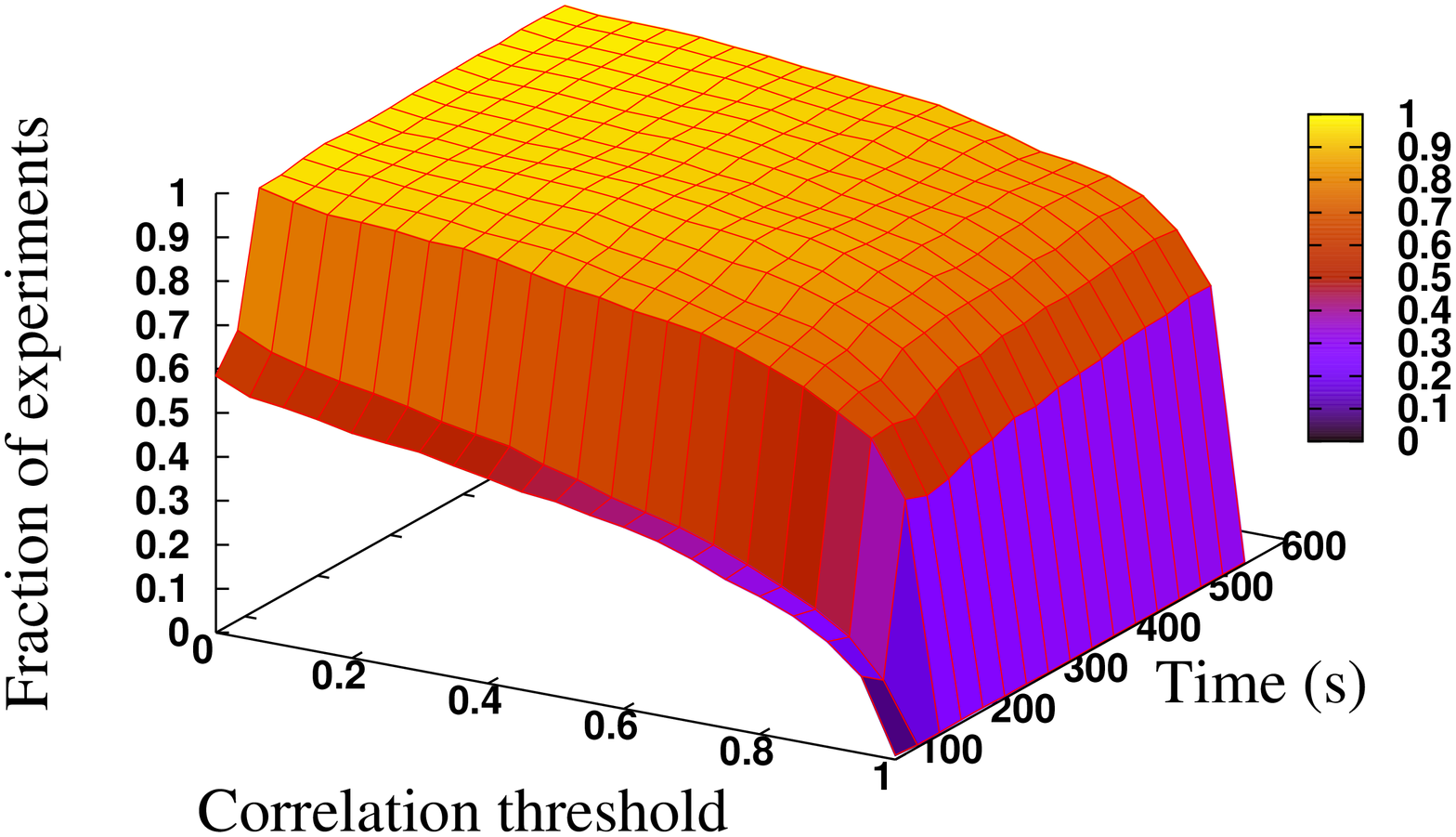}}}}
\hspace{0.1in}
\subfigure[c]{{{\includegraphics[width=1.5in,angle=270]{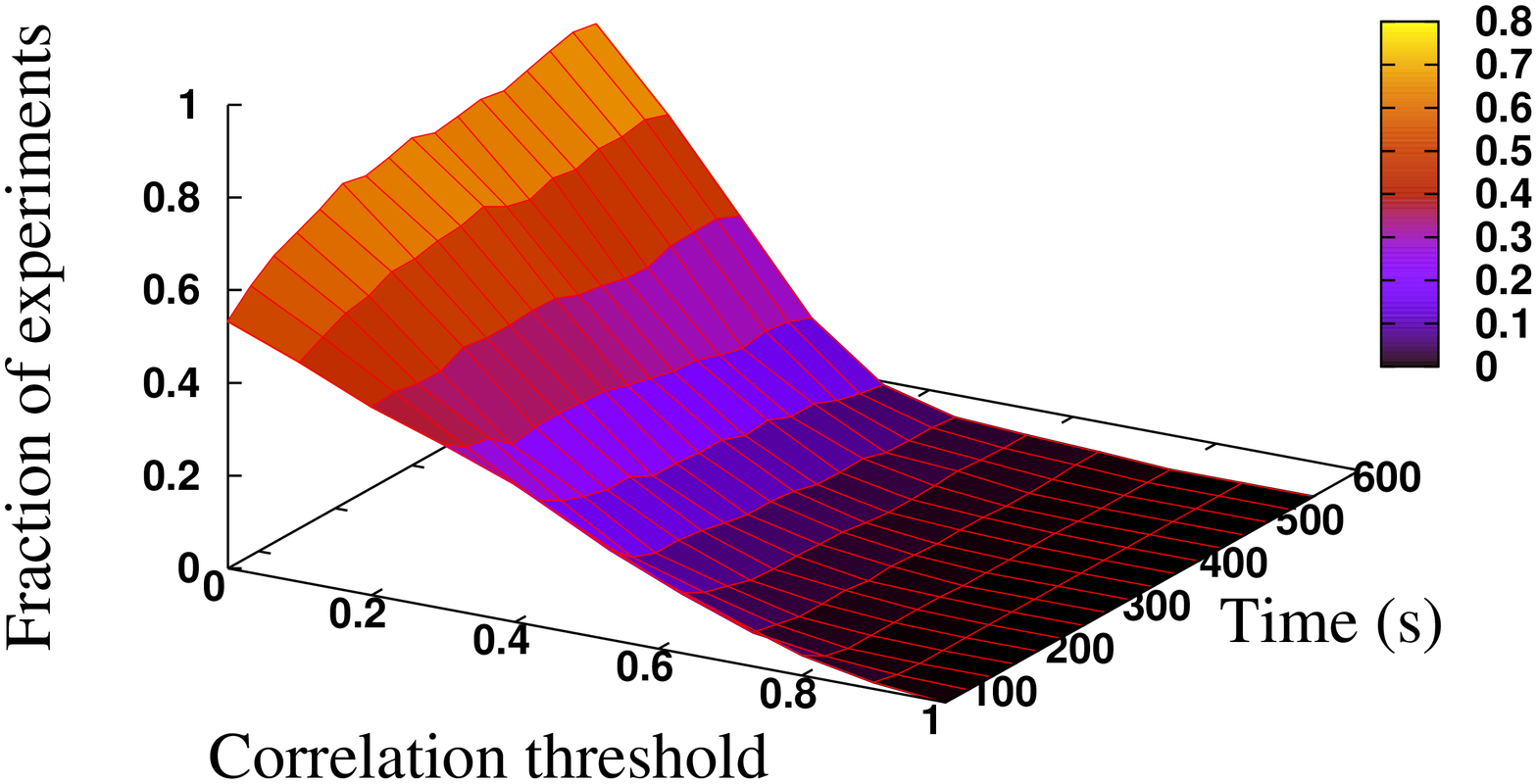}}}}
\hspace{0.1in}
\subfigure[c]{{{\includegraphics[width=1.5in,angle=270]{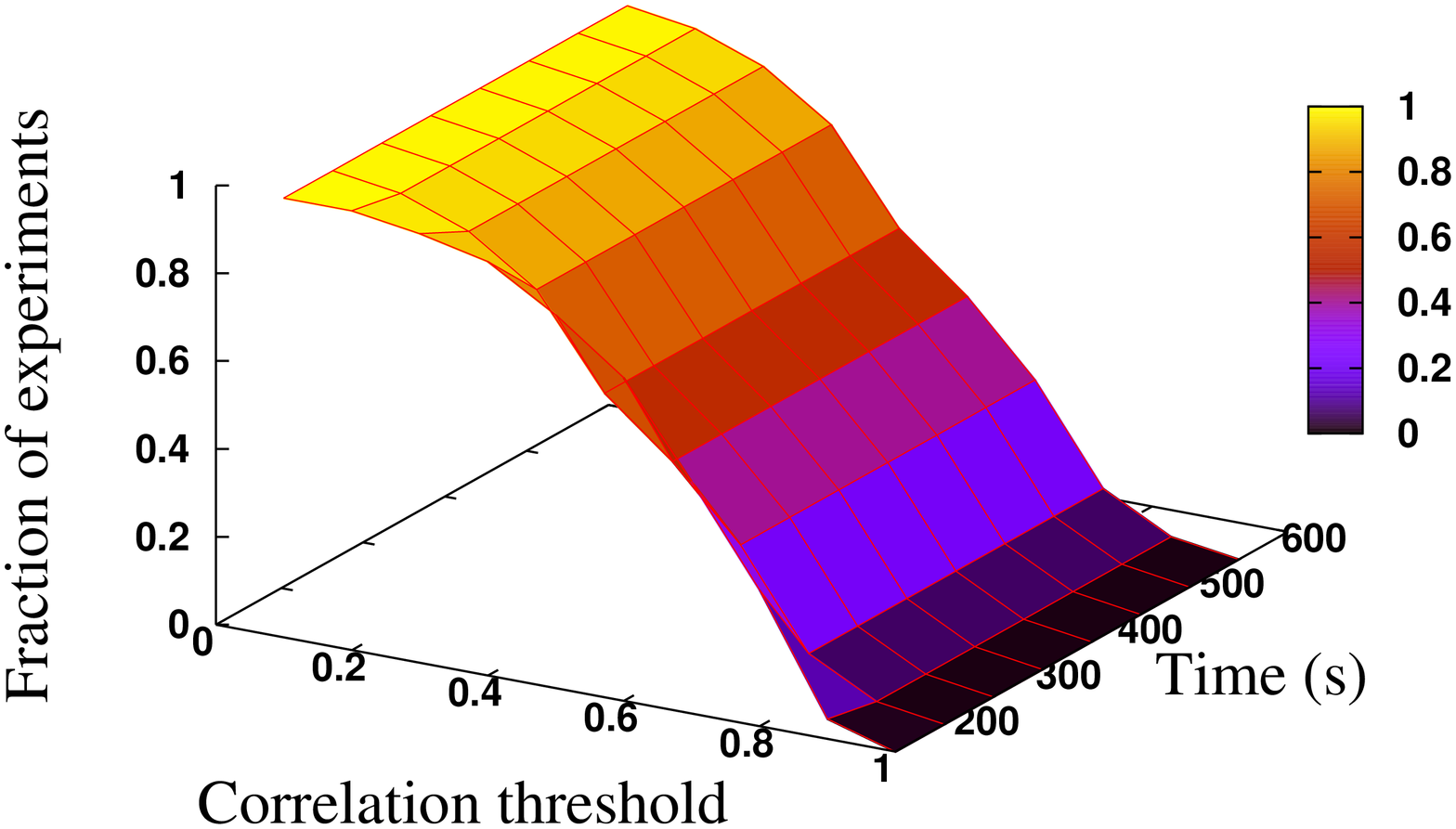}}}\label{fig:correlation-results:oc}}
\caption{
Effect of correlation threshold on detection rate: (a) \emph{All-Common},
(b)  \emph{None-Common}, (c)  \emph{One-Common-a}. 
}                                 
\label{fig:correlation-results}
\end{figure*}

\paragraphb{Results for  \emph{All-Common} (Figures \ref{fig:correlation-time}(a) and \ref{fig:correlation-results}(a)):}
We next present experimental results for the scenario where all three Tor 
relays are the same for both circuits. We performed $700$ 
runs of this experiment, between November and December 2010. 
We found that the throughput values for the two circuits were highly correlated\prateek{, measured using the Pearson product-moment correlation coefficient.} 
Figure~\ref{fig:correlation-time}(a) illustrates an instance of this scenario,  
where correlation is depicted as a function of time. 
We also compute the \nikita{95\%} confidence interval for correlation \matt{using Fisher's} 
Z transform~\cite{fisher-z}. We can see that after a time duration of 300 seconds, even the lower 
bound for the confidence interval is higher than $0.7$. Figure~\ref{fig:correlation-results}(a) depicts 
the fraction of experiments where the correlation value is greater than a threshold 
as a function of time. In about $92\%$ of the cases, the correlation was greater than 
a threshold value of $0.5$ after a time duration of $300$ seconds. Moreover, we can also see 
that correlation remains steady with time.

\paragraphb{Results for  \emph{None-Common} (Figures~\ref{fig:correlation-time}(a) and~\ref{fig:correlation-results}(b)):}
Next, we consider the scenario where two circuits are completely disjoint. We performed $532$ runs of this experiment 
in January 2011. In Figure~\ref{fig:correlation-time}(a), we can see that the correlation 
quickly approaches zero when the two circuits are comprised of disjoint relays. 
Figure~\ref{fig:correlation-results}(b) shows the full results for this scenario; we can see that in an overwhelming fraction of cases, the correlation 
quickly drops to zero---only $12$ instances out of $532$ had a correlation value greater than $0.5$ for a time interval of $300$ seconds (false 
positive percentage of only 2.2\%). In our investigation of false positives, we found that many of the
false positive instances occurred due to the geographic co-location of relays in the two circuits. For example, in \matt{four}
instances, the two circuits shared a relay that was part of the same subnet, like the \texttt{blutmagie} and \texttt{blutmagie4}
 relays. In \matt{one instance}, the first $16$ bits of the IPv4 address of two relays was the same (\texttt{ctor} and \texttt{cptnemo}, both 
located in Berlin), while in \matt{two} other instances, IP geolocation databases indicate geographic closeness within 
tens of miles.
However, geographic colocation could not explain the remaining \matt{five} instances.  

\paragraphb{Results for  \emph{One-Common-a} (Figures~\ref{fig:correlation-time}(b) and~\ref{fig:correlation-results}(c)):} 
Finally, we consider the scenario where two circuits have only a single common Tor relay. We build one-hop circuits 
through each of the relays in a client circuit. %
We performed $190$ runs of this experiment for this scenario between
November 2010 and February 2011.
Figure~\ref{fig:correlation-time}(b)
depicts the correlation as a function of time for two instances in this scenario. In scenario \emph{One-Common-a}, the common relay was the bottleneck 
relay for the full duration of 10 minutes, and we can see that the correlation value is similar to that of scenario \emph{All-Common}. On the other 
hand, in scenario \emph{One-Common-b}, the common relay was not the bottleneck relay, and we can see that the correlation is not statistically significant. 
When only a single relay is common between two circuits, we found that in many instances, correlation degrades over time, since the common 
relay may be a bottleneck only temporarily (not shown in Figure~\ref{fig:correlation-time}(b)). To address this issue, 
we modified the correlation analysis to consider \emph{windows} (intervals) of time; for a 
particular size of the window, we compute the maximum correlation in the data over all time intervals greater than 
the window size (if the communication duration is smaller than the window size, we compute correlation as before). \mn{Are you saying
for all intervals $(t_1,t_2)$ such that $t_2-t_1 > w$? I thought we'd only use intervals equal $w$ in length.}
We probed all three relays for a client circuit in this experiment, but since only one of them 
may be the bottleneck at a time, we computed the maximum correlation amongst the three probes for different 
communication duration values. 
Figure~\ref{fig:correlation-results}(c) depicts the fraction of experiments that have correlation greater than a threshold 
value for different communication durations using a window size of $200$ seconds. We can see that 
while correlation is not as high as the scenario where all three relays were common between two 
circuits, over $90\%$ and $80\%$ of the experiments still have correlation greater than $0.3$ and 
$0.4$ respectively after $300$ seconds of communication. Next, we will use this observation to 
identify Tor relays.  %

Based on these experiments, we conclude that high correlation between circuit throughput indicates the presence of common Tor relay(s) 
in the two circuits. Note that the converse is not necessarily true.  

\subsection{Identifying Tor Relays}
\label{sec:spechop}

\begin{figure}[!t]\centering
\includegraphics[width=2.5in]{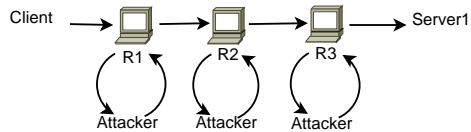}\\
\caption{
\prateek{Identifying Tor relays attack}: localizing which Tor relays are on a path by exploiting circuit throughput characteristics. Observe that the 
attack cost scales as $O(N)$.}\label{fig:deanonymize-relays-circuit2}
\end{figure}

\paragraphb{Threat model:} Next, we present an attack where we try to identify one or more Tor 
relays being used by a particular flow. We call this flow the \emph{target flow}. 
This flow can be any flow initiated by a Tor client over the 
Tor network to access a resource present at a particular server.   
\nikita{We assume that the attacker can observe the throughput of the target flow.
The attacker could have compromised the exit relay, the target web server,
or the ISP forwarding the data; note that the attacker does not need to
perform any modifications to the flow.}
The attacker tries to achieve its goal by probing 
different Tor relays; \nikita{importantly, these probes can be launched from different vantage points than the target server}. It builds one-hop circuits through these relays 
(Figure~\ref{fig:deanonymize-relays-circuit2}) and computes correlation\matt{s} between the throughput 
of the target flow and the \emph{probe flows}. If the throughput of a probe flow 
is highly correlated with that of the target flow, then the server can assume that both the flows 
are actually traversing \matt{a} common Tor relay. In this section, 
we show the success of the attacker at \prateek{identifying} Tor relays by 
running experiments over the live Tor network.

\paragraphb{Experimental setup:} In an ideal setting, 
the attacker will try to probe every relay in the Tor network and 
find \matt{the} correlation between the probe flows and the target 
flow. 
However, in order to reduce the amount of resources required to 
perform this attack, we \matt{used} a smaller set of $25$ Tor relays 
for constructing circuits. 
\prateek{To pick relays that are representative of the 
current Tor network, we selected relays with a probability proportional 
to their bandwidth using the Tor client. }     
We performed our experiments during two different time period
and used a different set of 25 relays in each period; we
will refer to them as   RELAY-SET-1 and RELAY-SET-2 (the list of Tor relays is available in 
\begin{techreport}
Appendix~\ref{sec:relays}.) 
\end{techreport}
\begin{ccs}
the full version of this paper~\cite{tech-report}).        
\end{ccs}

\begin{figure*}[t]
\centering
\subfigure[]{{{\includegraphics[width=1.7in,angle=0]{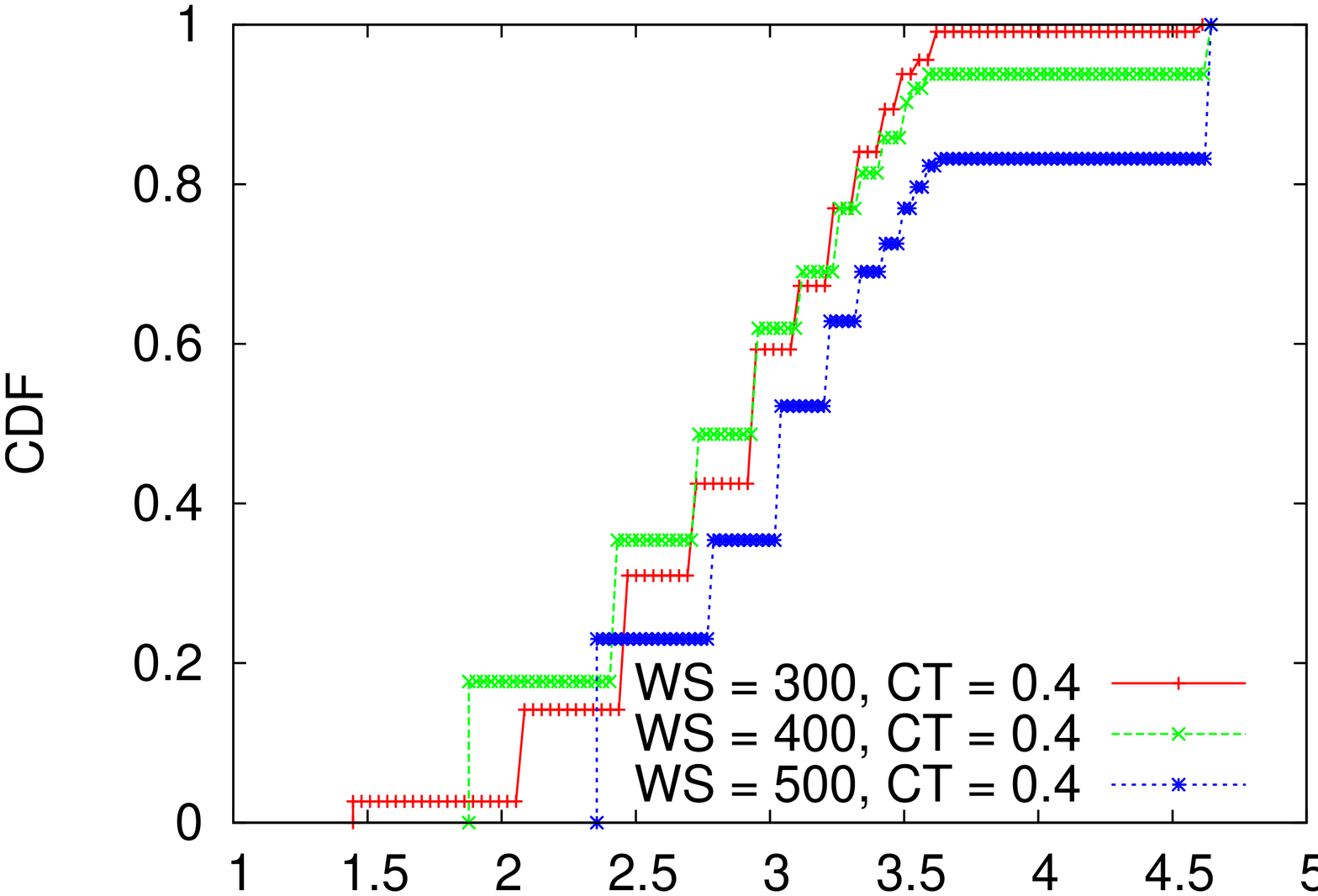}}}}
\hspace{0.4in}
\subfigure[]{{{\includegraphics[width=1.7in]{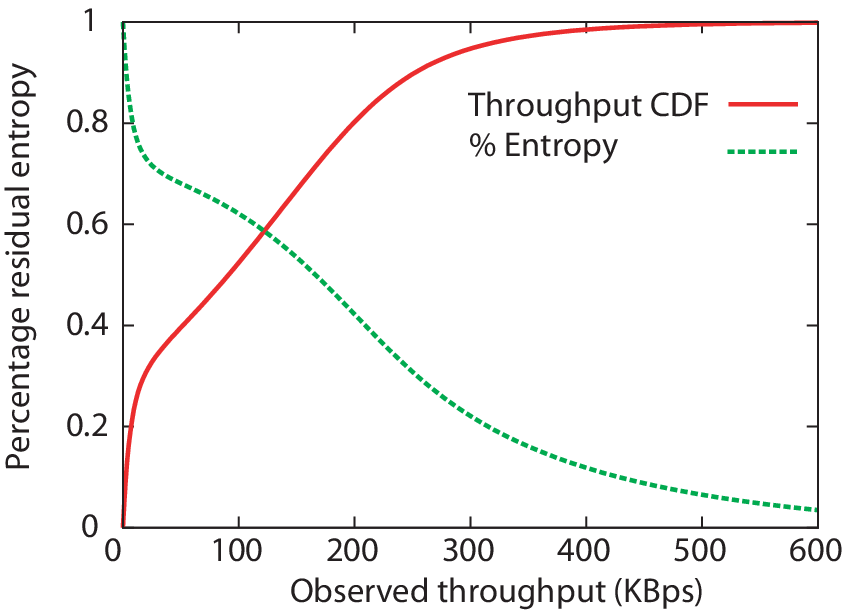}}}}
\hspace{0.4in}
\subfigure[]{{{\includegraphics[width=1.7in]{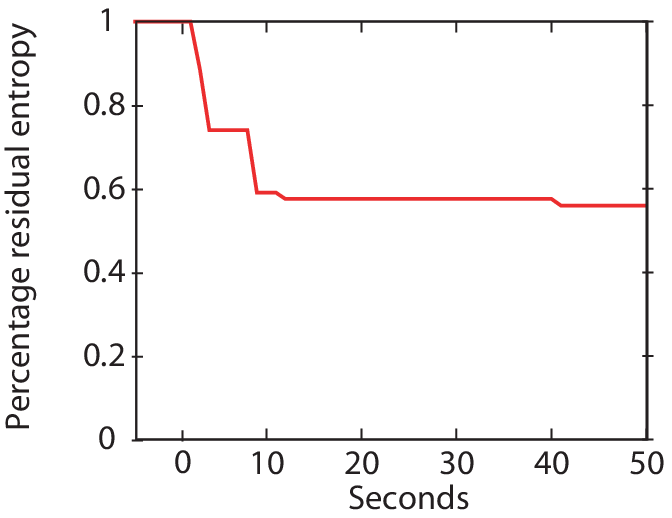}}}}
\caption{
(a) Entropy of the anonymity set obtained through our de-anonymizing relay attack. 
(b) Entropy loss and CDF of observed throughput.
(c) Decrease in entropy over %
six consecutive ten second intervals.
}
\label{fig:entropyfigs}
\end{figure*}

\paragraphb{Probe locations:} We used $25$ different Emulab~\cite{emulab} machines 
at the University of Utah %
to run the probe flows.\footnote{Note that we did not make use of
Emulab's network emulation infrastructure and ran experiments on the
production Tor network instead.} They were used to 
probe all the $25$ Tor relays simultaneously.      
\nikita{All of the probe flows originated from a single physical 
location, but we expect that their throughput was nevertheless 
shaped by the Tor overlay, rather than the access link: in a separate
test, we were able to achieve 70\,Mbps across the access link, far above the throughput of flows in the Tor network.}

\paragraphb{\matt{Single circuit experiments:}} 
In our first experiment, %
we set up a target flow between a client and a malicious server
over a circuit that was constructed using a set of $25$ Tor relays. %
We started all the probe flows simultaneously after the target flow ran 
for $50$ seconds. Each probe flow lasted for $600$ seconds, and we recorded the 
observed throughput every $0.1$ seconds. Overall, we performed $40$ measurements using 
RELAY-SET-1 in November 2010 and $150$ measurements using RELAY-SET-2 in January-February 2011. 
We use the entropy 
metric~\cite{diaz02,serjantov02} (with a uniform prior) to quantify the 
attacker's uncertainty about the relays in a circuit 
\begin{techreport}
(our computations are described in Appendix~\ref{sec:entropy}). 
\end{techreport}
\begin{ccs}
(our computations are described in technical report~\cite{tech-report}). 
\end{ccs}
Figure~\ref{fig:entropyfigs}(a) depicts the CDF of the bottleneck relay entropy for varying 
window sizes using a correlation threshold of $0.4$. We can see 
that while our attacks are not able to exactly identify the bottleneck relay, 
the bottleneck relay entropy is still drastically reduced. For example, 
using a window size of $300$ seconds, the entropy in half of the cases is 
less than $2.5$ bits (out of 4.6 bits).
We found similar results for correlation thresholds $0.3$ and $0.5$. %
\mn{Do we use a uniform prior here?  If so, this is wrong, but we should mention either way.}

\mn{Are we dealing with distributions of relays, or do we just have a number of relays that we eliminate?
If it's the latter, count should be enough.}
In addition to being able to reduce the %
\prateek{entropy of the bottleneck relay distribution},
we can also %
learn some information about the target flow's non-bottleneck relays. 
Our attack observation is that if the mean throughput of a one-hop probe through 
a relay is less than the mean throughput of the target flow, then that relay 
is unlikely to be part of the target flow circuit. 
The expected drop in entropy for a given throughput is shown in 
Figure~\ref{fig:entropyfigs}(b).  We also plot the CDF of Tor circuit throughput 
for comparison.
We can see that the circuits receiving a higher throughput have a lower entropy in the network. 
This is due to the presence of only a few high-throughput relays in the system; 
if high throughput is achieved, there are only a limited
number of relays capable of providing this service. The average value of \mn{The median would be better here.}
throughput observed in our experiment was 96\,KBps thus yielding an expected
instantaneous drop of about $18\%$ in the total entropy at a\nikita{ny} given point in the
experiment's run time.

We demonstrate the usefulness of this observation by eliminating relays with lower throughput than the target flow over a period of time. The results are plotted in Figure~\ref{fig:entropyfigs}(c). 
By using this simple technique, we were able to reduce the entropy in our sample space by 
roughly $40\%$ in the first ten seconds, and using a minute-long attack, we were able to reduce 
the entropy the bottleneck relay's sample space by roughly $45\%$ as fluctuating probe 
bandwidth measurements allowed more relays to be eliminated from consideration. 

\subsection{Identifying Guard Relays}
\label{sec:guards}

\begin{figure*}[t]
\centering
\subfigure[]{{{\includegraphics[width=1.5in,angle=270]{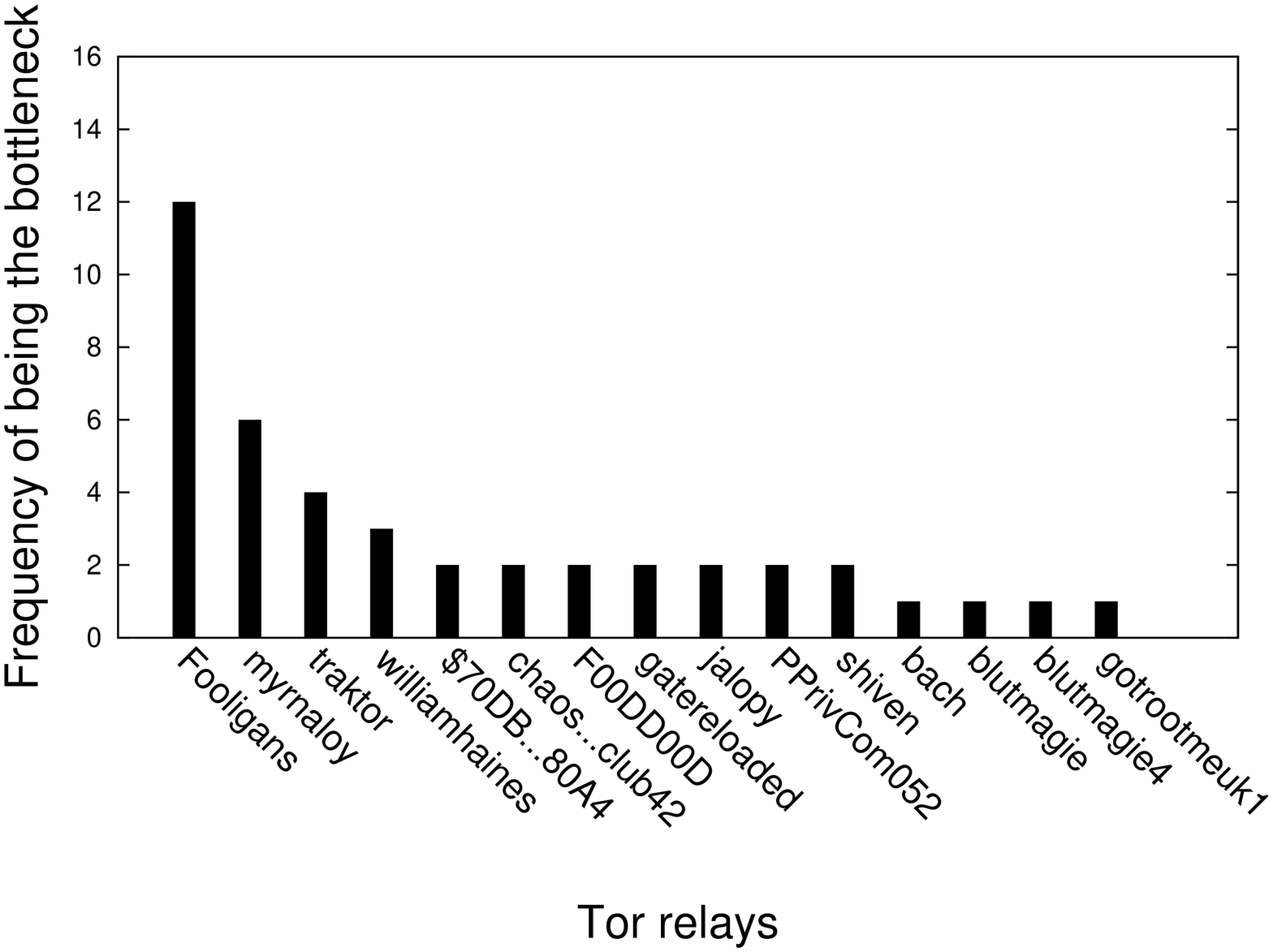}}}}
\subfigure[]{{{\includegraphics[width=1.5in,angle=270]{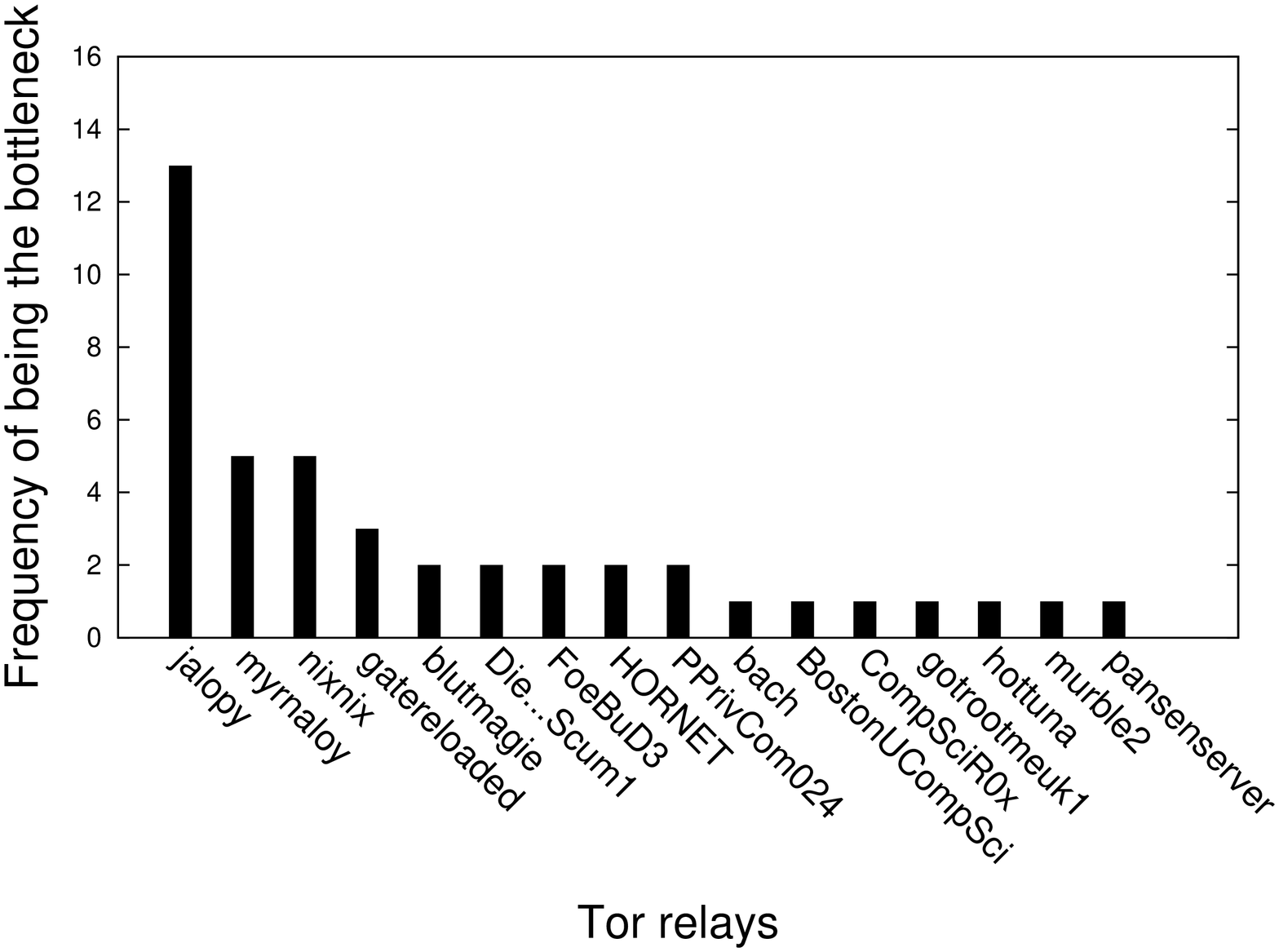}}}}
\subfigure[]{{{\includegraphics[width=1.5in,angle=270]{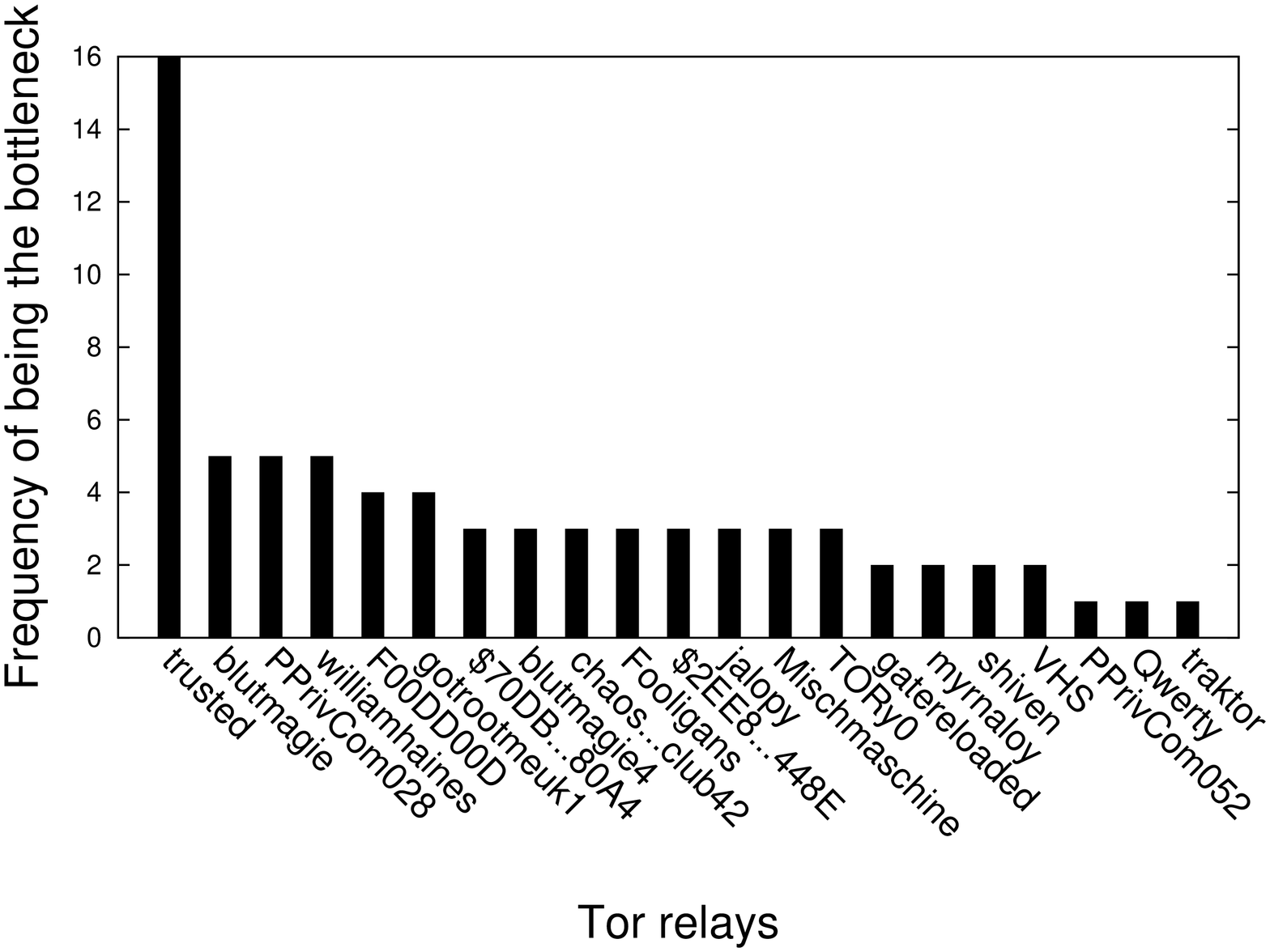}}}}
\vspace{-0.15in}
\caption{
Number of times a particular relay was \matt{detected as a} bottleneck,
with guard relay 
(a) \texttt{Fooligans}, with observed capacity $82$\,KBps,
(b) \texttt{jalopy}, with observed capacity $5.8$\,MBps,
(c) \texttt{trusted}, with observed capacity $17.4$\,MBps,
as of February 2011.
}
\label{fig:guard-attacks}
\end{figure*}

So far, we presented attacks that reduce the entropy of Tor relays in 
a single circuit. We now present attacks that can combine such 
probabilistic observations over multiple circuits, and \prateek{identify %
a client's set of guard relays}. \matt{Our 
observation} is that the guard relays are more likely to be 
present in the anonymity set of bottleneck relays for a client's circuit, 
as compared to other relays (since the probability of the guard relays
appearing in the client's circuit is much higher than other relays).
Thus, over multiple circuit reformulations, the set of relays 
that appear in the anonymity set for a client's circuit with the 
highest frequencies are likely to be the guard relays. 

\paragraphb{Experimental setup:} 
Here, in addition to restricting the client's Tor application in using relays 
from a selected set of $25$ relays, we also 
restricted its choice of guard relay. 
With the help of the configuration options {\texttt{EntryNodes}, \texttt{StrictEntryNodes}, 
\texttt{UseEntryGuards}, and \texttt{NumEntryGuards}}, we configured the client's Tor application 
to always use a particular relay as the guard relay. 
The intuition behind this experiment is to analyze the vulnerability of 
using a fixed guard relay; i.e., we want to observe whether use of a 
fixed guard relay allows the attacker to combine probabilistic observations 
over multiple circuits and %
\prateek{identify the guard relay}. 
Similar to the experiments for \prateek{identifying %
 Tor relays}, we used $25$ different 
machines to do the probing where the target flow ran for $700$ seconds and the probe flows 
ran for $600$ seconds (within the lifetime of the target flow).
We performed three experiments for this scenario, where we chose the guard relays 
to be \texttt{Fooligans}, \texttt{jalopy}, and \texttt{trusted}, respectively.  We performed 
$50$, $20$, and $50$ runs for the three experiments respectively, between November 2010 
and February 2011. Note that we considered 
these three relays since we wanted to study the effectiveness of our attack against guard 
relays with very different capacities. \texttt{Fooligans}, \texttt{jalopy}, and \texttt{trusted} 
have a reported capacity of 82\,KBps, 5.8\,MBps, and 17.4\,MBps, respectively (as of 
February 2011).

\paragraphb{Results:}
Figure~\ref{fig:guard-attacks} show the number of times a particular relay 
was selected in the anonymity set for the bottleneck relay used by the target flow. We used 
a correlation threshold of $0.4$, and set the window size to be $400$. 
In Figure~\ref{fig:guard-attacks}(a), we can see that the relay with the highest 
frequency of being in the anonymity set of bottleneck relays is \texttt{Fooligans}, which 
is indeed the guard relay for the client. In fact, in all three experiments, our attack 
was able to \prateek{infer %
the identity of the guard relays}, since they appeared in 
the anonymity set with the highest frequencies.

Observe that accounting for the fact that Tor clients 
use $3$ guard relays by default (as compared to one in our experiments) 
would reduce the frequency of them being detected 
as a bottleneck relay by a factor of $3$.
However, note that our results are conservative since we selected relays from a 
limited pool of only $25$ Tor relays. The current Tor network size is about $2\,000$ 
relays; application of this attack on the full Tor network would result in 
orders of magnitude reduction in the selection probability of relays which are 
not the guard relay, correspondingly reducing their frequency of being detected 
as a bottleneck relay. Thus in the full Tor network, 
even with $3$ guard relays, we can expect the frequency of the guards being detected 
as the bottleneck relay to be much higher than a random relay in the network. 

Finally, we note that our attack was able to \prateek{identify %
 the guard relay} 
irrespective of its observed capacity. As the Tor network is heavily 
used by a large number of users, even high capacity relays become 
congested and cannot ensure enough throughput to all the flows. This causes 
them to affect flows' throughput in a way that can be detected through 
our one-hop %
\prateek{relay identification attack}. As the guard relay has a direct 
connection with the client, the severity of this attack is 
much higher than an attack that identifies the middle relays. %

\subsection{De-anonymizing Tor Relays Offering \\ Location Hidden Service}
\label{sec:lhs}

In this section, we attempt to identify Tor relays that also offer 
location hidden services. The main purpose of a location hidden 
service is to remain anonymous while serving its clients. However, 
we show that when a Tor relay offers a location hidden service, 
it becomes susceptible to attacks. The observation that enables 
de-anonymization of location hidden servers is similar to our attack 
on guard relays, in that the location hidden service is likely 
to have the highest frequency of being part of the anonymity set 
of bottleneck relays for circuits connecting to that location 
hidden service. 
In order to assess the effectiveness of this attack, we set up our 
own Tor relay and advertised its existence to the Tor network. 
We also configured this machine to offer a location 
hidden service. %
We setup a target flow from a client machine to the location hidden service, and 
measured the flow throughput. While this target flow was active, we probed our 
Tor relay using a one-hop circuit as we did in Section~\ref{sec:spechop}. 
We ran this experiment $100$ times and computed the correlation between the 
throughput of the target flow and the probe flows. 
Using a correlation threshold of $0.4$, and window sizes of $300$ and $400$ seconds, 
we found that $15\%$ and $9\%$ of the target flows and the corresponding probe flows 
were highly correlated.  
Relating this observation with the observations from Section~\ref{sec:spechop}, we can 
conclude that our Tor relay will be placed in the anonymity set 
for bottleneck relays with a much higher probability than any other relay. 
Moreover, as there are usually $8$ different relays/nodes including the server itself 
(the other seven relays comprise of three relays each for the client's circuit as well 
as the server's circuit, and a common meeting point for these circuits, called 
the \emph{rendezvous} relay) 
in the path between the client and the server, each of them can become the bottleneck 
with $12.5\%$ probability. Results from our experiment closely matches this 
value by identifying our Tor relay with roughly the same probability (9\%-15\%).

\subsection{Applicability to Interactive Traffic}
\label{sec:interactive}

\begin{figure}[]
\centering
\includegraphics[width=1.4in,angle=270]{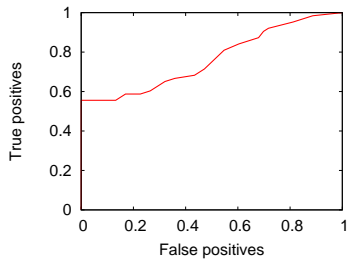}
\caption{Attack on interactive traffic: ROC curve.}
\label{fig:Interactive-ROC}
\end{figure}

So far, we used bulk traffic in our experiments. 
We now investigate %
the accuracy of our attacks with interactive traffic 
(characterized by bursts of traffic
followed by periods of inactivity). 
To evaluate our attacks, we first built a model for interactive traffic by 
collecting real packet level traces from a fast Tor guard relay at various times 
from April 24 through April 27. 
\prateek{To preserve user privacy, %
we first used Mittal et al.'s~\cite{mittal:hotnets09} anonymizing framework to process the pcap traces; 
for each observed packet, the output of the framework was an anonymized client IP address, packet timestamp, 
and packet length. We performed the above data anonymization and minimization steps on the guard relay itself. 
The original pcap traces were deleted immediately afterwards\footnote{In future work, we will enhance the anonymizing 
framework to perform the data minimization and anonymization steps in memory, in order to further minimize risk in case 
the guard relay is compromised during the data processing phase.}. Next, we analyzed the properties of the processed data. 
} 
In total over the three days we observed $19\,478$ flows %
with $17\,911$ of these %
flows containing between one to ten packets. %
We surmise that these correspond to either preemptive circuits which were never actually used 
by Tor or failed circuit creation attempts. 
For the remaining flows, we break them up into ten-minute intervals and then 
calculate the fraction of 5 second sub-intervals where any traffic
was observed.  We next classify them into three categories based on the fraction of time they were observed 
sending data.
We found 424 flows that transferred traffic for more than 90\% of the time (non-interactive 
file downloads), 347 flows with between 50\% and 90\% utilization (interactive flows with moderate utilization), 
and 796 flows with less than 50\% utilization (interactive flows with low utilization). 
We select only the flows comprising interactive traffic with moderate utilization, 
and use the distribution of their average burst sizes and gap times as a model for interactive traffic in
our experiments. \prateek{After computing the distribution of burst sizes and gap times, all intermediate 
data from previous steps was discarded to maintain the privacy of Tor users.}
Next, we performed two experiments over the live Tor network, with 100 measurements each for the Scenarios All-Common and None-Common. 
In both experiments, we set up a bulk transfer flow (the attacker) and an interactive flow (the victim).
For the interactive transfer, the TCP server was configured to send data to the victim in bursts where 
each burst was followed by a gap period. %
Both the burst size and the gap times were sampled from the observed distribution computed using real data.
For both experiments, we computed the correlation in throughput of the attacker flow and the victim flow for 
those time intervals where the victim flow was active. We are interested in the fraction of measurements 
where the correlation was greater than a threshold. This metric corresponds to true positives for the 
Scenario All-Common, and false positives for the Scenario None-Common. The correlation threshold determines 
the trade off between the true positives and the false positives. Figure~\ref{fig:Interactive-ROC} 
depicts the ROC curve for our experiments. We found that for an estimated 
false positive rate of 0 (95\% confidence interval is less than 3.5\%), the true positive rate is greater 
than 0.5. Our analysis demonstrates the applicability of our attacks on interactive traffic, though the 
accuracy in this case is lower than that of bulk traffic. 

 \section{Stream Fingerprinting}
\label{sec:stream-throughput}
\label{sec:stream-attacks}

Circuits in Tor typically carry multiple \emph{streams}, which correspond to individual TCP connections.
By default, all streams created by a client
within a 10 minute period are carried over the same circuit.\footnote{\nikita{This is governed by the 
\texttt{Max\_Circuit\_Dirtiness} parameter in Tor.}}  
Tor uses a different scheduling mechanism to share bandwidth between
streams than it does between circuits, providing an opportunity for new
attacks \nikita{that correlate streams}. In particular, we
will show that packet scheduling of %
streams that use the same circuit 
creates special characteristics that can
easily be recognized by a completely passive attacker; this can be used to link the streams as belonging to the same user and thus 
compromise some of the unlinkability protections provided by Tor.\footnote{It should be noted that exit relays can already perform such \nikita{linking}; our attacks,
however, do not require the compromise of any relays in the Tor network.}

\begin{figure}[]\centering
\includegraphics[width=2.5in]{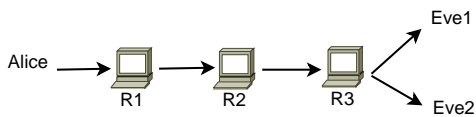}\\
\caption{Stream fingerprinting attack: breaking unlinkability by exploiting stream throughput characteristics.}\label{fig:unlinkability-stream}
\end{figure}

\subsection{Stream Throughput at Different Time Scales}

\begin{figure*}[t]
\centering
\subfigure[]{{{\includegraphics[width=2.1in]{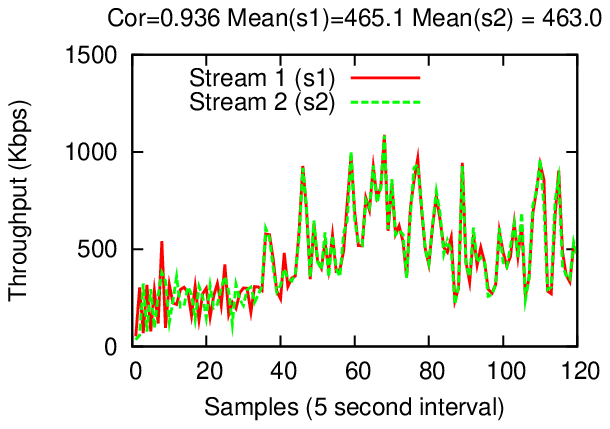}}}}
\hspace{0.2in}
\subfigure[]{{{\includegraphics[width=2.1in]{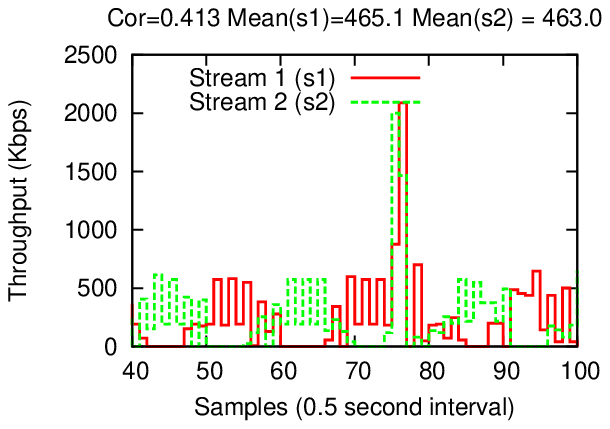}}}}
\hspace{0.2in}
\subfigure[]{{{\includegraphics[width=2.1in]{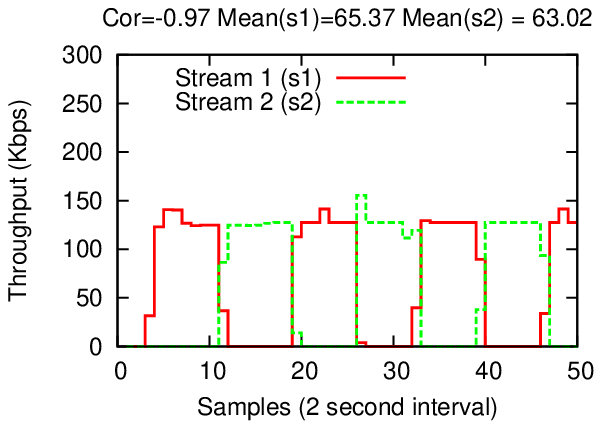}}}}
\caption{
(a) Positive correlation between streams.
(b) Batching effect when sampling every 0.5 seconds.
(c) Batching effect when sampling every 2 seconds.
}
\label{fig:batching}
\end{figure*}

Consider the scenario in Figure~\ref{fig:unlinkability-stream}: Alice is communicating simultaneously with 
Eve1 and Eve2. Alice's Tor client multiplexes both streams over the same circuit, in this case consisting of 
Tor relays $R1,R2,R3$. Eve1 and Eve2 can see that both their connections are coming from the same exit node, 
and would like to \nikita{learn} if they are both communicating with the same person.  
We would expect that, since the streams share the same path through the network, they should
exhibit similar throughput characteristics. %
Indeed, we find a very strong correlation between such streams; 
Figure~\ref{fig:batching}(a) 
shows
the results of one experiment, with a correlation of 0.93.  However, as shown in Section~\ref{sec:attacks},
a high correlation is common among unrelated circuits that share the same bottleneck node, e.g., if $R3$
is the bottleneck for both streams.

However, there is a second effect that can be used to distinguish streams that share one or more common relays. %
Tor implements stream-level throttling within circuits. 
This end-to-end flow control works as follows: each stream 
is initially allowed to send $500$ of Tor's 512-byte cells 
 (also called the \emph{packaging window}). Upon successful reception 
of packets, the packaging window is incremented in blocks of $50$ cells. Observe that 
if a stream is waiting for the packaging window to be incremented, then another stream 
that is being multiplexed on the same circuit can hog the circuit (\emph{batching effect}) for a small 
time interval.                            

While the streams share the throughput fairly, flow- and congestion-control mechanisms in Tor result in batching 
behavior, where one stream may monopolize the circuit for some period of time, while the other streams lie idle.  We can see this  mutually exclusive use of the circuit by two streams when we sample the throughput
at smaller time scales, as can be seen in 
Figures~\ref{fig:batching}(b) and~\ref{fig:batching}(c).
We note that the length of the batches seems 
to be a multiple of 25\,KB, corresponding to 50 of Tor's 512-byte cell, matching our intuition that
50-cell packaging window increments are responsible for the batching effect. This also means that the time scales necessary to observe this effect will vary depending on the throughput of the flow:
in 
Figure~\ref{fig:batching}(b), 
the mean throughput value is 450\,Kbps and we observe the behavior sampling every $0.5$ seconds, 
whereas in 
Figure~\ref{fig:batching}(c), 
the mean throughput value is 66\,Kbps, and we observe the behavior sampling every $2$ seconds.

Based on these observations, we propose the following algorithm to infer whether the streams are coming from 
the same client:
\begin{enumerate*}
    \item We first check for strong correlation at macro-level time scales, as done in previous sections. 
    \item We define our micro-level time scale to be $\frac{25\,\text{KB}}{\text{Throughput(KBps)}}$
     \item We check for the batching effect at micro-level timescales.  Namely, we compute the 
	fraction of time when only one stream is active out of the total time that either stream is active, which shows the degree of ``mutual exclusivity'' between the two streams.
\end{enumerate*}

\subsection{Results}
To see how well the algorithm works, we performed an experiment in which a single
client connects to two servers under our control, located in two different geographical 
locations, 
and simultaneously downloads data. Since the connections were 
made simultaneously, the two streams were multiplexed over the same Tor circuit.  
We performed a total of 225 runs between November 2010 and February 2011. 
To check for false positives in our attack, we repeated the experiment with downloads by two different clients using
separate circuits that have a common exit relay. %
We performed 225 iterations of this experiment in April 2011.

\begin{figure}[]\centering
\includegraphics[width=2.1in]{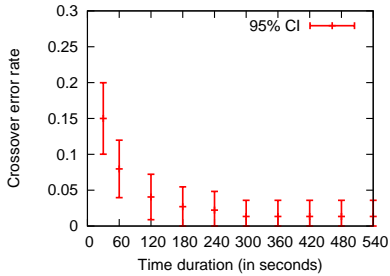}\\
\caption{Stream linkability attack: crossover error rate as a function of time.}\label{fig:stream-cor}
\end{figure}

Recall that the tradeoff between false positives 
and false negatives is governed by the choice of the threshold values for correlation and mutual exclusiveness. 
As the threshold values are increased, the false positives decrease, but the false negatives increase.
We plot the crossover error rate as a function of the communication duration in Figure~\ref{fig:stream-cor}. We can 
see that as the time duration is increased, the accuracy of our attack quickly improves. Even at a small duration 
of 120 seconds, our mean crossover error rate is less than $5\%$, which gets reduced to $1.3\%$ beyond $300$ seconds (with a 95\% confidence
interval of less than 4\%).
Hopper et al.~\cite{hopper:ccs07} analyzed a similar attack scenario, where latency information was used to determine if
the same client is communicating with two servers. However, the crossover error rate in their attack was 17\% even when 
the communication duration was 10 minutes. %
Note that throughput and latency provide orthogonal sources of information, and it may be interesting to combine 
our attacks with latency attacks to further improve accuracy.
Finally, we also study the scenario where client Alice is communicating simultaneously on 3 streams, but 
only two of the destination endpoints are colluding. %
We found that the crossover error rate in this case was higher than the previous scenario, but still 
less than $8\%$ when the communication duration is $300$ seconds.

\section{Discussion}
\label{sec:discussion}

We discuss some issues and ramifications of our attacks:

\paragraphb{Countermeasures:}
We now consider some possible ways to defend Tor users against our attacks.  
First, the circuit throughput fingerprinting
attack  relies on the existence of a shared bottleneck relay along the two
paths.  One countermeasure might be to use better load balancing within the Tor
network.  For example, preferring paths through higher capacity relays may make
it more likely that the bottleneck is outside the Tor network.  While it may be
difficult to completely eliminate all bottlenecks within the Tor network, it
may be possible to avoid making more sensitive relays as bottlenecks, for
example by setting up paths such that guard relays are not the bottlenecks.  In
addition, multipath forwarding has been shown to give more even load balance in
conventional networks~\cite{trump}.  Using multipath forwarding could reduce
the number of bottlenecks by evenly distributing traffic over multiple paths,
as well as ``average out'' short-term variations in throughput that arise from
any single path.  Also, since our attacks take some time to reliably associate
the circuits, endpoints may periodically cycle through a small set of circuits,
to reduce the amount of time they spend sharing a bottleneck with a malicious
circuit.  Another approach might be to change the effects that the bottleneck
relay has on traffic. For example, relays may intentionally use an uneven rate
allocation across flows, or inject spurious cross-traffic, to decouple
throughput correlations in circuits they carry.  Next, the stream linkability
attack is based on Tor's intra-stream scheduling mechanism.  To avoid this
attack, Tor users may wish to perform heuristics at the client side to prevent
circuit re-use for sensitive flows.  In addition, better stream multiplexing
algorithms~\cite{DefenestraTor} (e.g., algorithms that share bandwidth on a finer granularity) could
eliminate the batching effects that we observed.

\paragraphb{Alternative schemes to measure throughput:}
Our current attacks measure throughput by simply forwarding TCP traffic through
the Tor relays.  However, much work has been done \matt{in the networking
community} in the area of throughput probing, and it may be possible to leverage
these techniques for improved performance.  For example, instead of just
monitoring flow throughput, techniques exist to monitor total {\em capacity} of
the bottleneck link, by sending packets with precise
timings~\cite{hl+05,pathchar}. In situations where these techniques can be
applied, correlation may be simplified, by revealing how much total capacity
there is in nodes participating in the underlying channel.  In addition, some
of these techniques can measure throughput within short bursts of packets (e.g.,
packet pairs and packet trains~\cite{hl+05,pathchar}), which may improve
correlation for sessions of short duration.

\begin{techreport}
\paragraphb{Alternatives to probing:}
Our study indicates that throughput information is sensitive, and can be used
to reveal private information about clients.  This calls into question
proposals to make observed bandwidth information public.  For example, there
has been work~\cite{eigenspeed} on sharing bandwidth information between peers,
to improve load balancing.  In the context of Tor, the Tor directory service
reveals information about the bandwidth of relays.  There have also been
measurement studies on Tor~\cite{mikeperry,karsten} that have conducted
bandwidth measurements of Tor and released these results into the public
domain, which may reveal information about Tor users.
\end{techreport}

\paragraphb{Resource requirements:} %
Our primary attacks
are {\em stealthy} and %
do not require the attacker to congest Tor relays. %
Thus, the only cost incurred by an attacker is to create circuits 
to narrow the set of relays present in a circuit. 
To get a loose upper bound on the attacker's resource requirements, we
note that the average throughput through a Tor relay from our study of 2,104 relays 
(Figure~\ref{fig:capacity-throughput-relays}) is $226$\,KBps. Consider the very extreme case where the
attacker simultaneously constructs one-hop circuits through each of the Tor
relays in the entire network. The attacker would need to send $464.36$\,MBps
($1632.51$\,GB per hour) of traffic. While this amount is higher than typical edge
node bandwidth, the attacker may be able to acquire such bandwidth by 
using botnet hosts, or a hosted infrastructure such as Amazon EC2. Using EC2
pricing of \$0.12 per GB, the attacker could send traffic for one hour with a cost of 
\$196 (volume pricing for bandwidth in EC2 would reduce this cost for
longer monitoring periods).

\paragraphb{Impact of bulk traffic throttling:} 
Recently, Tor implemented 
a mechanism \prateek{that enables relay operators} to throttle bulk traffic connections using a token bucket 
rate limiting system on traffic coming from non-relays. The rate limiting 
takes into account the  burst rate and the long term throughput rate of a connection. 
The mechanism was first implemented in Tor-0.2.2.15-alpha, and thus did 
not impact our initial experiments conducted throughout November of 2010. 
However, as new versions of Tor were deployed throughout early 2011, we noticed that 
some of the attacker's probe circuits were getting throttled after 
roughly 300 seconds of high rate transfer down to a set limit of about 20\,KBps. 
Note that the attacker can easily overcome the effects of throttling on its probe 
circuits by opening a new probe circuit when the current circuit begins to be throttled. 

\paragraphb{Ethics:} Given that throughput information can leak information
about clients, we conducted our study with some caution.  In our
experiments, we only \prateek{attack} %
the circuits created by our own 
experimental tool to avoid collecting information about non-participating
clients.  We also avoided launching large-scale attacks; we focus most of our
experiments on only 25 Tor relays. All throughput data we collected was stored with strong
encryption on machines behind a firewalled network. 
\prateek{While analyzing traffic on a guard relay, we only recorded data pertaining to the 
distribution of burst sizes and gap times for the interactive traffic model (data-minimization). All other collected data (including outputs 
of various intermediate steps) was discarded to maintain the privacy of Tor users.}
%
%
%

%
%
%
%
%
%
%
%
%
%
%
%
%
%

%

%
%
%
%
%
%
%
%
%
%
%
%

%
\section{Concluding Remarks}
\label{sec:conclusion}

To the best of our knowledge, our work comprises the first study of 
{\em throughput attacks} on anonymity systems.
We presented attacks to identify Tor relays used by an existing connection,
as well as techniques to identify the relationship between two streams,
solely by monitoring their throughput. %
This calls into question, existing proposals to make observed bandwidth information 
public~\cite{eigenspeed,karsten,mikeperry}. %
Our techniques are stealthy (cannot be detected by Tor users/relays) and low cost (do not 
require Tor relays to be congested). Moreover, our attacks use threat models and resource 
requirements that differ significantly from attacks using other flow features. Overall, our 
study highlights the complexity of designing anonymous communication systems.

\section*{Acknowledgments}

We are grateful to Roger Dingledine for helpful discussions about Tor and for his guidance in 
shepherding our paper. We would like to thank Eugene Vasserman for suggesting the mechanism to 
probe Tor relays. This paper benefitted from discussions 
with George Danezis, Amir Houmansadr, Qiyan Wang, Giang Nyugen, Sonia Jahid and Xun Gong. 
Finally, we would like to thank the anonymous reviewers for their invaluable feedback on 
earlier drafts of this paper. 
This work was supported in part by NSF CNS 08-31488, 09-53655, 10-40391 and an International Fulbright S\&T Fellowship.

{
\small
\bibliographystyle{acm}
\bibliography{hatswitch}
}

\begin{techreport}
\newpage
\appendix

\section{Relays Used in Experiments}
\label{sec:relays}

\begin{table}[h]
\footnotesize
\centering
\begin{tabular}{|c|c|}
\hline
RELAY-SET-1 & RELAY-SET-2  \\
\hline
\hline
bach & \$70DB...80A4 \\ %
BelieveInGood & agate \\
blutmagie & bach \\
BostonUCompSci & blutmagie \\
CompSciR0x & blutmagie4 \\
customer0648f5 & CB3ROB \\
DieYouRebelScum1 & chaoscomputerclub42 \\
fejk4 & F00DD00D \\
FoeBuD3 & Fooligans \\
gatereloaded & \$2EE8...448E \\ %
gotrootmeuk1 & gam \\
HORNET & gatereloaded \\
hottuna & gotrootmeuk1 \\
jalopy & jalopy \\
mitternacht & Mischmaschine \\
murble2 & myrnaloy \\
myrnaloy & PPrivCom028 \\
nixnix & PPrivCom052 \\
pansenserver & Qwerty \\
Pasquino & shiven \\
PeaShooter & TORy0 \\
PPrivCom024 & traktor \\
Roo8Peik & trusted \\
SunnySmile & VHS \\
torforpresident & williamhaines \\
\hline
\end{tabular}
\caption{Tor relays used in the probe experiments.}
\label{tbl:probe-exp-relays}
\end{table}

\section{Entropy Metric}
\label{sec:entropy}

This metric considers the
\emph{distribution} of potential relays in a circuit, as computed by attackers, 
and computes its entropy:
\begin{equation}
H(C) = -\sum_c p_c \log_2 p_c %
\end{equation}
\noindent where $p_c$ is the probability that the relay $c$ was part of the          circuit.  
Under some observation $o$, we can compute the probability distribution given $o$    and compute the 
corresponding entropy $H(C|o)$.  To model the entropy of the system as a whole, we   compute a 
weighted average of the entropy for each observation (including the null             observation): 
\begin{eqnarray}
 H(C|O) = \sum_o P(o) H(C|o) %
\end{eqnarray}
\noindent where $P(o)$ is the probability of the observation $o$ occurring, 
and $O$ is the set of all observations. This is also known as the conditional 
entropy of $C$ based on observing $O$. Now, let $S$ be the set of relays for which 
the one-hop probes have high correlation with the target flow. Also, let $P(R_S)$ 
denote the probability that the correct relay is included in the set $S$. 
The attacker can compute $H(C|S)$ as follows: 
\begin{equation}
 H(C|S) = P(R_S) \log_2\frac{|S|}{P(R_S)} + (1-P(R_S)) \log_2\frac{|N-S|}{1-P(R_S)} 
\end{equation}

\end{techreport}

\end{document}